\documentclass[journal]{IEEEtran}
\usepackage{cite}

\usepackage{hyperref}

\ifCLASSINFOpdf
   \usepackage[pdftex]{graphicx}
\else
\fi
\usepackage{amsmath}
\usepackage{url}

\newcommand\Ksunit{$\mathrm{m}^{1/3}.\mathrm{s}^{-1}$}

\usepackage{color}
\usepackage{caption}
\usepackage{subcaption}

\usepackage{amssymb}                  
\usepackage{amsmath}
\usepackage{multirow}
\usepackage{graphicx}
\usepackage{xfrac}
\usepackage{bm}

\begin{document}
%
\title{Improvement of Flood Extent Representation with Remote Sensing Data and Data Assimilation}
%
%
%

\author{Thanh~Huy~Nguyen,
        Sophie~Ricci,
        Christophe~Fatras,
        Andrea~Piacentini,
        Anthéa~Delmotte,
        Emeric~Lavergne,
        and Peter~Kettig
\thanks{Manuscript received September 16, 2021; revised November 30, 2021; accepted January 16, 2022. This work was supported by CNES and CERFACS. \textit{(Corresponding author:
Thanh Huy Nguyen.)}}
\thanks{T.H. Nguyen, S. Ricci, and A. Piacentini are with the Centre Européen de Recherche Avancées et de Formation en Calcul Scientifique (CERFACS), Toulouse,
31057, France (e-mail: thnguyen@cerfacs.fr; ricci@cerfacs.fr; piacentini.palm@gmail.com).}
\thanks{A. Delmotte was with the Centre Européen de Recherche Avancées et de Formation en Calcul Scientifique (CERFACS), Toulouse,
31057, France (e-mail: delmotte@cerfacs.fr).}
\thanks{C. Fatras and E. Lavergne are with the Collecte Localisation Satellites (CLS), Toulouse, 31000, France (e-mail: cfatras@groupcls.com; elavergne@groupcls.com). }
\thanks{P. Kettig is with the Centre National d'Etudes Spatiales, Toulouse, 31000, France (e-mail: peter.kettig@cnes.fr).}
}

%
%

\markboth{Journal of \LaTeX\ Class Files,~Vol.~14, No.~8, August~2015}%
{Shell \MakeLowercase{\textit{et al.}}: Bare Demo of IEEEtran.cls for IEEE Journals}
%



\maketitle

\begin{abstract} 
Flood simulation and forecast capability have been greatly improved thanks to advances in data assimilation. Such an approach combines in-situ gauge measurements with numerical hydrodynamic models to correct the hydraulic states and reduce the uncertainties in the model parameters. However, these methods depend strongly on the availability and quality of observations, thus necessitating other data sources to improve the flood simulation and forecast performances. Using Sentinel-1 images, a flood extent mapping method was carried out by applying a Random Forest algorithm trained on past flood events using manually delineated flood maps. The study area concerns a 50-km reach of the Garonne Marmandaise catchment. Two recent flood events are simulated in analysis and forecast modes, with a +24h lead time. This study demonstrates the merits of using SAR-derived flood extent maps to validate and improve the forecast results based on hydrodynamic numerical models with Telemac2D-EnKF. Quantitative 1D and 2D metrics were computed to assess water level time-series and flood extents between the simulations and observations. It was shown that the free run experiment without DA under-estimates flooding. On the other hand, the validation of DA results with respect to independent SAR-derived flood extent allows to diagnose a model-observation bias that leads to over-flooding. Once this bias is taken into account, DA provides a sequential correction of area-based friction coefficients and inflow discharge, yielding a better flood extent representation. This study paves the way towards a reliable solution for flood forecasting over poorly gauged catchments, thanks to available remote sensing datasets.
\end{abstract}


%
\IEEEpeerreviewmaketitle

\section{Introduction}
%
%
%
%

\subsection{Flood monitoring}
Flooding causes major economic losses and a serious threat to human life and subsistence. It is one of the most devastating natural hazards that our society must adapt to worldwide, especially as the severity and the occurrence of flood events tend to intensify with climate changes \cite{IPCC2021}, \cite{IPCC2021SPM}.
At national scale, flooding accounts for the major part of the extreme hazards recorded over the last five years in France and concern about seventeen million people in 2017 \cite{odry2017comparison}.
Several international initiatives have joined efforts in research, observation and computational science programs dedicated to flood monitoring, in order to provide governments, decision support systems and insurance companies with improved simulation and observation solutions. 
Such efforts from the community of environmental remote-sensing (RS) monitoring led, for instance, to the International Charter on Space and Major Disasters\footnote{\url{www.disasterscharter.org}}, a unified system for space data acquisition and delivery to those affected by disasters, such as flooding, via its member space agencies. 
With the rapidly increasing volume of data from space, Earth observation is now at the core of international programs such as Copernicus, in particular the Emergency Management Service (Copernicus EMS with Mapping and Global/European Flood Awareness System services) that provide actors involved in the management of natural disasters with relevant satellite RS data. The Space Climate Observatory (SCO) coordinates space agencies and international organizations to assess and monitor the consequences of climate change from observations and numerical models, especially at local scales. 
In this regard, the Flood Detection, Alert and rapid Mapping project (labelled as FloodDAM) has been supported by the SCO initiatives. 
It aims at developing pre-operational tools to enable quick responses in various flood-prone areas while improving the resolution, reactivity and predictive capability \cite{kettig}. 
In France, the SCHAPI (Service Central d'Hydrom\'et\'eorologie et d'Appui {\`a} la Pr\'evision des Inondations) and French Flood Forecast Services (SPCs) are in charge of monitoring and forecasting water level and discharge over 22,000 km of rivers. 
They produce a twice-daily vigilance color-coded risk map available online for governmental authorities and the general public\footnote{\url{www.vigicrues.gouv.fr}}. To create these risk maps, they rely on hydrodynamic numerical models and in-situ measurements and also investigate the use of RS data.

\subsection{Improving hydrodynamic models with data assimilation}
Hydrodynamic models use the amount of water entering the river system to compute water level and velocity in the river network, and when the storage capacity of the river is exceeded, in the flood plain. These models are used to predict river water surface elevation and velocity from which flood risk can be assessed for lead-times that range from a couple of hours to several days. They solve the shallow water equations (SWE) derived from the free surface Navier-Stokes equations \cite{swe2d}~\cite{navierstokes}. However, these numerical codes are imperfect as uncertainties inherently existing in the models and in the inputs (model parameters, boundary conditions, geometry) translate into uncertainties in the outputs. The performance of the hydrodynamic model is limited by the amount and quality of available data  (e.g., \cite{Bates1997, Bates2004,Wood2016,Horritt2000,Werner2005}). Model parameters such as friction coefficients are usually calibrated for significant flood events with respect to observational data. As a result, the model can only be calibrated and validated as finely as the available data allows it, stressing out the need for a time and space densified observing network \cite{mirouze2019impact}. 
Otherwise, hydrodynamic models remain imperfect, especially with respect to the dynamics of the flood plains, and thus should be further validated and improved.

Data assimilation (DA) aims at estimating the optimal state of a model by sequentially combining the model and the observations while taking into account their respective uncertainties \cite{ide1997unified,kalnay2003atmospheric,asch2016data}. While historically employed in meteorology and oceanography, DA is now commonly used in hydrology with hydrodynamic models to improve discharge and water level forecasting as well as risk of marine submersion and flooding \cite{liu2012advancing}. Sequential DA here allows to reduce the most significant sources of uncertainties in the hydrodynamic models related to friction coefficients and input forcing data, which constitute the control vector. In this regard, the Ensemble Kalman Filter (EnKF) \cite{evensen1994sequential} is favored for model parameter and forcing correction as it stochastically estimates the covariances between the control vector and the observation errors \cite{BARTHELEMY2017210}.
In the context of DA, the use of RS data allows to overcome the limits due to the lack and decline of in-situ river gauge stations \cite{ad2001global}. It especially allows for validation of DA results in flood plains with independent observations. Jafarzadegan \textit{et al.} \cite{Jafarzadegan2019} illustrates how the use of the maximum inundation maps (delineated after a flood event) allows to validate a dual state-parameter DA strategy of in-situ gauges where correlation between observation errors are accounted for. RS-derived data here are presented as a complement of in-situ data that is sparse or absent in the case of ungauged catchments.

\subsection{Remote sensing flood extent observation for calibration and/or validation of hydrodynamic models}\label{ssec:rsdata} 
The use of RS products in the context of flood risk management presents a great opportunity to improve the ability to monitor and forecast flooding as stated by \cite{Schumann2009}.
Indeed, in the recent years, Synthetic Aperture Radar (SAR) data have increasingly become one of the most efficient ways to map and monitor flood extents in near-real time over large areas, due to their all-weather day-and-night imaging capabilities \cite{hostache2009water}. 
Water bodies and flooded areas typically exhibit low backscatter intensity on SAR images as most of the incidence radar pulses are specularly reflected away upon arrival \cite{henderson1998principles}. 
Thus, the detection of these areas on SAR images is relatively straightforward, with several exceptions such as built environments and vegetated areas that yield backscattering similar to that of permanent water bodies and flooded areas.
As a matter of fact, many research works have leveraged SAR data for flood extent mapping and flood depth estimation.
Chini et al. \cite{chini2017hierarchical} proposed a change detection approach applied on SAR images called hierarchical split-based approach. It aims at discriminating the two classes, namely wet pixels (for flooded areas) and dry pixels in SAR images by estimating their respective backscatter intensity distribution.
Cian et al. \cite{cian2018flood} proposed a two-step flood depth estimation using different RS datasets. Firstly, the flood extents were delineated on SAR images using Normalized Difference Flood Index \cite{cian2018normalized}. Such an index is computed based on statistical analysis of backscatter values from two SAR multi-temporal image stacks, one from a reference time period and another involving flood events. Then, the flood depth was estimated from the flood extents thanks to a LiDAR-based Digital Elevation Model (DEM). The merits of RS data were also demonstrated for the automatic estimation of flood event duration using Sentinel-1, Sentinel-2 and Landsat-8 data \cite{rattich2020automatic}.

The increasing availability of highly spatially distributed RS observations of flood extent and water levels offer new opportunities for investigation and analysis (e.g., \cite{Bates2004,Schumann2009}). 
The possibility of using SAR imagery data for the validation and calibration of two-dimensional (2D) hydraulic models was first  highlighted by Jung et al. \cite{Jung2012}. Since then, the increasing amount of RS data and the advances in Machine Learning algorithms dedicated to water detection, have enabled a great number of research work dedicated to hydrology and hydraulics models calibration and validation for real-time forecasting. The combination of RS data with local hydrodynamic models has thus been greatly studied in the literature as it allows to overcome the limitations of both incomplete and uncertain sources of knowledge on the river and flood plain dynamics.
The last available comprehensive review by Grimaldi et al. \cite{grimaldi2016remote} provides an analysis on the use of coarse-, medium- and high-resolution RS observations of flood extent and water level to improve the accuracy of hydraulic models for flood forecasting. It points out that RS data should be used as a complement data source---but not as an alternative---to the in-situ data in order to calibrate, validate, and constraint the hydraulic models. This stems from their low precision and acquisition frequency \cite{grimaldi2016remote}. Indeed, compared to in-situ data, RS data provide useful flood extent and flood edge information at a large coverage, usually covering the whole considered catchment, but they are much sparser in terms of frequency. In addition, uncertainty exists in flood extent mapping from RS observations, e.g. SAR images, which originates from both the input images and the classification algorithm itself. As a matter of fact, classification overall accuracy of flooded areas varies considerably and only in rare cases exceeds 90\% \cite{Schumann2012}. An updated review from Dasgupta et al. \cite{Dasgupta2021review} provides the state-of-the-art on the assimilation of Earth Observation data with hydraulic models for the purpose of improved flood inundation forecasting.

A common use of RS-derived flood extent for hydrodynamic model calibration and validation requires to retrieve water surface elevation (WSE) or river width information with complementary DEM data. Mason et al. \cite{mason2012automatic}, Gustarini et al. \cite{giustarini2011assimilating} proposed to take advantage of the flood edges derived from SAR images as observational information. As such, flood edges were extracted from SAR images and then integrated with an available DEM to derive the WSE on the flood plain. Once the coherence of the hydraulic state is achieved, this 2D information is then compared with and/or assimilated to the WSE simulated by one-dimensional (1D) hydrodynamic model with a particle filter data assimilation algorithm sequentially updating the hydraulic state. This strategy demonstrated great capability in re-analysis and forecast mode. Yet the definition of the control vector involving only the hydraulic state was stated as a limitation when larger rivers are considered and the correction of the upstream boundary condition should be considered to improve forecasts. A similar approach was proposed by Scarpino et al. \cite{scarpino2018multitemporal} which leverages multitemporal COSMO-SkyMed SAR images and a DEM to derive flood depth maps to calibrate channel and flood plain friction coefficients, in order to achieve flood dynamics monitoring in a flat area with complex topography. \cite{grimaldi2018effective} illustrated how RS-derived river width can be combined with a limited number of river depth measurements and empirical functions to estimate simplified river geometries for 2D hydrodynamic models dedicated to inundation.

Flood probability maps have also been estimated by a Bayesian approach, namely a Particle Filter (PF),  applied on SAR images, and subsequently assimilated into a particle filter-based data assimilation framework \cite{hostache2018near}. This approach saves the burden to retrieve water depth from SAR images (which requires well-known topographical information). The particles representing each individual hydraulic forecast are forced with perturbed rainfall inputs and weighted with respect to the disagreement between the forecast and the observed flood probability (inferred from SAR images) at the assimilation times. Several follow-up papers focus on the improvement of this strategy. For instance, \cite{DiMauro2021} enhances how a tempering coefficient is used to reduce the degeneracy of the Particle Filter and how this results in improved forecasts. \cite{Dasgupta2020} formulates the likelihood function within the Particle Filter based on mutual information which accounts for spatial uncertainty correlation and also leads to an improved forecast. The sensitivity to the observing network characteristics was investigated in the framework of synthetical twin experiments where probability floop maps generated from a reference run and backscatter values issued from distributions of flood and non-flood typical classes \cite{Dasgupta2021network}.

Flood extents derived from SAR images based on the processing of the open water backscatter probability density function proposed in \cite{matgen2011towards,Giustarini2013,Chini2016}, were used in \cite{Wood2016} to estimate friction and bank-full depth with the  dynamic identifiability analysis (DYNIA) algorithm \cite{wagener2003towards} applied to the LISFLOOD model \cite{neal2012simple}. It was shown that this method does not require the expression of SAR-derived water level and allows for a dual calibration that improves as the number of SAR images used in the algorithm increases. 
Cooper et al. \cite{cooper2019observation} proposed a new observation operator that directly uses backscatter values from SAR images as observations in order to bypass the flood edge identification or flood probability estimation processes. However, this approach has only concerned synthetical SAR images since it relies on the hypothesis that SAR images must yield distinct distributions of wet and dry backscatter values. This may not be the case for real data when the mean backscatter values of wet and dry pixels are close. It should be noted that the comparison of non-hydrometric observations, e.g. flood edge locations, flood probability measures, derived from SAR images with the hydrodynamic model outputs is not straightforward and require the development of appropriate observation operators. Lastly, it should be emphasized that the combination between SAR images and hydrodynamic models depends strongly on the precision of flood extent detection and mapping techniques, as noted in \cite{DiMauro2021} in areas where backscatter values are not impacted by the appearance of floodwater (dense vegetation, urban areas), as well as on the quality of the numerical model (e.g. ability to initiate with and represent permanent water surfaces). Drawbacks on both sources of information on the flood extent should thus be taken into account in the comparison and validation step; this is investigated in the present work.

\subsection{Objective and outline}
This paper highlights the merits of using RS-derived flood extents to validate and improve a hydrodynamic numerical model with DA. In-situ water level observations are assimilated in a 2D model over the Garonne River in order to sequentially correct the friction and inflow discharge. The Telemac software\footnote{\url{www.opentelemac.org}} is used to simulate flooding for two major events that occurred over the Garonne Marmandaise in December 2019 and January-February 2021. 
The flood extents were derived from Sentinel-1 images by Machine Learning algorithm (Random Forest), and then compared with the flood extents simulated by Telemac. Statistical metrics are computed to evaluate the model performance with respect to these independent data that are not assimilated in the present work. We illustrate how, during significant flood events,  the use of spatial data overcomes the limits of river-gauge only validation process.

The remainder of the paper is organized as follows. Section~\ref{sect:MaterialData} gathers the method, material, and data used in this study. The hydrodynamic solver and models of the Garonne catchment over which the study is carried out are presented, respectively, in \autoref{subsect:SWE} and \autoref{subsect:T2D}. 
Subsection~\ref{subsect:algoEnKF} presents the Ensemble Kalman filter algorithm and its implementation. Employed RS data and the Random Forest algorithm carried out to retrieve flood extents from Sentinel-1 SAR imagery data are detailed in \autoref{subsect:floodml}. Finally, the metrics used in the paper are presented in \autoref{subsect:metrics}, they allow to assess the performance of the simulation and assimilation with respect to both in-situ and RS data as well as evaluating the quality of the ensembles in analysis and forecast mode.
Experimental results are presented in Section~\ref{sect:Results}, the merits of using RS data for simulation evaluation are highlighted. The improvement of flood peak simulation and forecast thanks to in-situ data assimilation are quantified. Discussion and limitations for this study are presented in Section~\ref{sect:Discussions}. Conclusions, limitations and perspectives are given in Section~\ref{sect:ConcPers}.

\section{Method}\label{sect:MaterialData}

\subsection{Shallow water equations in Telemac2D}\label{subsect:SWE}
    
The non-conservative form of the SWE are written in terms of  water level (denoted by $H$ [m]) and horizontal components of velocity (denoted by $u$ and $v$ [m.s\textsuperscript{-1}]). They express mass and momentum conservation averaged in the vertical dimension, assuming that (\textit{i}) vertical pressure gradients are hydrostatic, (\textit{ii}) horizontal pressure gradients are due to displacement of the free surface, and that (\textit{iii}) horizontal length scale is significantly greater than the vertical scale. The SWE read:
\begin{equation}
\begin{split}
		\label{eq:SWE1}
		\frac{\partial H}{\partial t} + \frac{\partial}{\partial x} \left( Hu \right) + \frac{\partial}{\partial y} \left( Hv \right) = 0 
\end{split}
\end{equation}

\begin{equation}
\begin{split}
		\label{eq:SWE2}
        \frac{\partial u}{\partial t} + u \frac{\partial u}{\partial x} + v \frac{\partial u}{\partial y}  = - g \frac{\partial Z}{\partial x} + F_x +
				 \frac{1}{H} div \left(H{\nu}_{e} \overrightarrow{grad} \left(u \right) \right)
\end{split}      
\end{equation}

\begin{equation}
\begin{split}
		\label{eq:SWE3}
        \frac {\partial v}  {\partial t} + u \frac {\partial v}  {\partial x} + v \frac {\partial v}  {\partial y} = - g \frac {\partial Z}  {\partial y} + F_y +
				  \frac {1}  {H} div \left( H {\nu} _ {e} \overrightarrow{grad} \left(v \right) \right) 
\end{split}      
\end{equation}
\\
where  $Z$ [m NGF69] is the water surface elevation ($H=Z-Z_b$ with $Z_b$ [m NGF69] being the bottom elevation) and $\nu _e$ [m\textsuperscript{2}.s\textsuperscript{-1}] is the water diffusion coefficient. 
$t$ stands for time and $g$ is the gravitational acceleration constant.
\textit{div} and \textit{$\overrightarrow{grad}$} are respectively the divergence and gradient operators.

In addition, $F_x$ and $F_y$ [m.s\textsuperscript{-2}] are the horizontal components of external forces (friction, wind and atmospheric forces), defined as follows:
 \begin{equation}
\begin{split}
 {F}_{x}  = & - \frac{g}{K_s^2} \frac{u\sqrt{u^2+v^2}}{{H}^{\frac{4}{3}}} 
				 - \frac{1}{\rho_{w}} \frac{\partial{P}_{atm}}{\partial x}  \\
				& + \frac{1}{H} \frac{{\rho}_{air}}{{\rho}_{w}} \; \; {C_d} \; \; {U}_{w,x} \sqrt {{U}_{w,x}^{2} + {U}_{w,y}^{2}}
\end{split}      
\end{equation}
 \begin{equation}
\begin{split}
{F} _ {y}   = & - \frac {g}  {{K_s^2}} \frac {v \sqrt{u^2+v^2}}  {{H} ^ {\frac{4}  {3}}} 
        - \frac {1}  {\rho _ {w}} \frac {\partial {P} _ {atm}}  {\partial y} \\
        & + \frac {1}  {H} \frac {{\rho} _ {air}}  {{\rho} _ {w}} \; \; {C_d} \; \; {U} _ {w , y} \sqrt {{U} _ {w , x} ^ {2} + {U} _ {w , y} ^ {2}}
\end{split}      
\end{equation}				
\\
where $\rho_w$/$\rho_{air}$ [kg.m\textsuperscript{-3}] is the water/air density ratio,  $P_{atm}$ [Pa] is the atmospheric pressure,  $U_{w,x}$ and $U_{w,y}$  [m.s\textsuperscript{-1}] are the horizontal wind velocity components, $C_d$ [-] is the wind drag coefficient that relates the free surface wind to the shear stress, and lastly, $K_s$ [m\textsuperscript{${\frac {1} {3}}$}.s\textsuperscript{-1}] is the river bed and flood plain friction coefficient, using the Strickler formulation \cite{gauckler1867etudes}.
 
In order to solve Eq. \eqref{eq:SWE1}-\eqref{eq:SWE3}, initial conditions $ \lbrace H(x,y,t=0)=H_0(x,y)$; $u(x,y,t=0)=u_0(x,y)$; $v(x,y,t=0)=v_0(x,y) \rbrace $ are provided, and boundary conditions (BC) are described with a time-dependent hydrogram upstream and a rating curve downstream. The Strickler coefficient is prescribed uniformly over defined subdomains, and calibrated according to the observing network. As aforementioned, in the present study, the SWE are solved with the parallel numerical solver Telemac2D, henceforth denoted by T2D, with an explicit first-order time integration scheme, a finite element scheme and an iterative conjugate gradient method \cite{hervouet2007hydrodynamics}. 
 
\begin{figure*}[t]
\centering
    \includegraphics[width=0.9\linewidth]{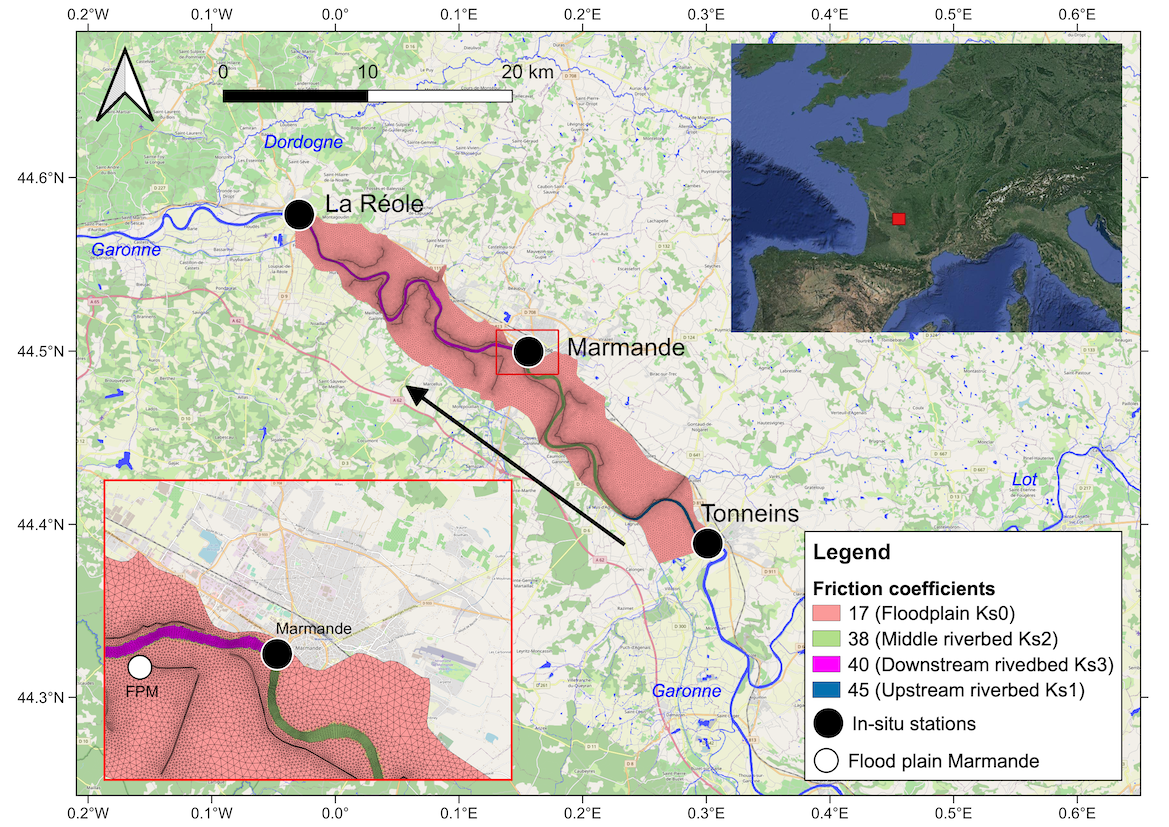}
     \caption{Study area of the Garonne River (southwest France, as shown in the upper-right corner inset figure) 50-km reach between Tonneins (upstream) and La Réole (downstream). 
     The black arrow indicates the flow direction. The black solid circles represent the in-situ Vigicrue observing stations.
     The inset figure at lower-left corner magnifies the area around Marmande.
     The white solid circle indicates a diagnosis location on the flood plain near Marmande (FPM).
     The friction coefficient $K_s$ is uniform over 4 zones: upstream, middle, and downstream river bed and flood plain.
     Background image: Map data \copyright OpenStreetMap contributors and available from \url{https://www.openstreetmap.org}.}
     \label{fig:study_area}
\end{figure*}

\subsection{The T2D model for the {\it Garonne Marmandaise}}
\label{subsect:T2D}
The study area extends over a 50-km reach of the Garonne River (southwest France) between Tonneins (upstream), downstream of the confluence with the river Lot, and La Réole (downstream) (\autoref{fig:study_area}). This part of the valley is identified as an area at high flood risk. Since the 19th century, it has been equipped with infrastructures to protect the Garonne flood plain from flooding events such as the historic flood of 1875. A system of longitudinal dykes and weirs was progressively constructed to protect flood plains and organize submersion and flood retention areas. Observing stations operated by the Vigicrue network are located at Tonneins, Marmande and La Réole (indicated as black solid circles on \autoref{fig:study_area}) and provide water level measurements every 15 minutes. The white circle indicates a chosen location in the flood plain near Marmande; it will be noted FPM in the following and used for diagnosis only. A T2D model was developed and calibrated over this reach  \cite{besnard2011comparaison}. It is built on a triangular unstructured mesh is used, with an increased mesh resolution around the dykes and in the river bed. 

The boundary conditions are prescribed upstream at Tonneins, and downstream at la Réole. The local rating curve at Tonneins (established from a limited number of water level-discharge measurements) converts the observed water levels into a discharge that is applied over the entire upstream interface (both river bed and flood plain boundary cells). This modeling strategy was implemented by Electricité de France (EDF) R\&D. While it has the merits to allow for a proper cold start of the model for any inflow discharge value (i.e. no dry grid cell at the initial time step on the upstream boundary), it also leads to the over-flooding of the upstream first meander. The downstream boundary condition at La Réole is described with a local rating curve \cite{Horritt2002}. 

\begin{figure}[h]
\centering
\includegraphics[width=0.95\linewidth]{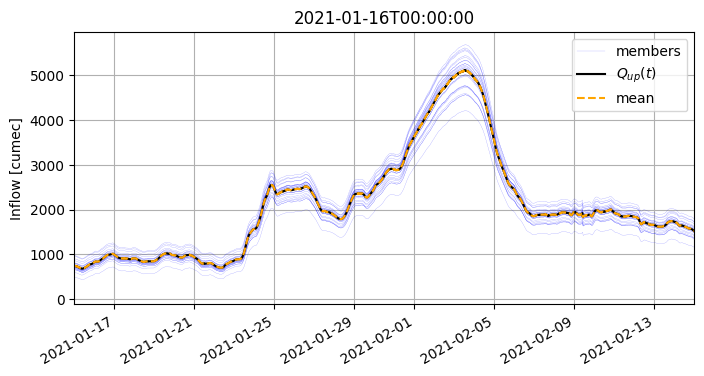}
\caption{Hydrographs used in the first EnKF cycle during the 2021 event. The black curve represents the \textquotedblleft true\textquotedblright~value of the inflow, whereas the $Q'_{up}$ used for the 24 members are depicted by light blue curves and the orange dashed curve stands for their average.}
\label{fig:q_up}
\end{figure}

The friction coefficient $K_s$ is defined over four areas as shown in \autoref{fig:study_area}. The friction coefficient values result from a calibration procedure over a set of non-overflowing events and are set respectively equal to: $K_{s_1}=45$, $K_{s_2}=38$ and $K_{s_3}=40$ \Ksunit~over the upstream, middle and downstream part of the river bed and $K_{s_0}=17$ \Ksunit~over the flood plain. It should be noted that the limited number of in-situ measurements restricts the spatial description and calibration of the friction in the river bed and the flood plain, leading to  uniform values within these areas and discontinuous values between these areas. In this study, both epistemic (due to the lack of correct parameter setting) and aleatory (due to the lack of true physical values) uncertainties are considered; 
as such, errors in water level and discharge are associated with errors in the friction or errors in the upstream BC. The $K_s$ coefficients setting is indeed prone to uncertainty related to the zoning assumption, the calibration procedure and the set of calibration events. This uncertainty is more significant in the flood plains where no observing station is available.
The probability density function (PDF) for the Strickler coefficients is assumed to follow a normal distribution with mean and standard deviation set accordingly to the calibration process and expert knowledge. The limited number of in-situ gauge measurements also yields errors in the upstream inflow as the expression of the inflow relies on the use of a rating curve, usually involves extrapolation for high flows \cite{Dibaldassarre2009}. In order to account for uncertainties in the upstream time-dependent discharge $Q_{up}(t)$ while limiting the dimension of the uncertain input space, the perturbation to BC is applied via a parametric formulation that allows for a multiplicative, an additive and a time-shift error, as proposed by \cite{ricci2011correction}: 
 \begin{equation}
Q_{up}^{\prime}(t) = a \times Q_{up}(t-c) + b
\label{eq:q(t)}
\end{equation}
where $(a,b,c) \in \mathbb{R}^3$, and their PDF are gaussian, centered at their default values $a=1, b=0, c=0$ such that $Q_{up}^{\prime}(t) = Q_{up}(t) $. The characteristics of the friction- and inflow-related uncertainty PDFs are given in \autoref{tab:PDFs}. \autoref{fig:q_up} depicts the ensemble of hydrographs used in the first EnKF cycle during the 2021 event, where the mean values of parametric coefficients from Eq. \eqref{eq:q(t)} are: $\bar{a}=1, \bar{b}=0, \bar{c}=0$.

\begin{table}[h] 
\centering
\caption{PDF of uncertain input variables related to friction coefficients $(K_{s_{[0:3]}})$ and inflow discharge corrective coefficients $(a, b, c)$.}
\begin{tabular}{cccc}
\hline
Variable & Calibrated/ & Standard & 95\% confidence \\
& default values ${\bf x}_0$ & deviation $\sigma_{\bf x}$ & interval \\\hline
$K_{s_0}$ & 17 & 0.85 & 17 $\pm$ 1.67 \\\hline
$K_{s_1}$ & 45 & 2.25 & 45 $\pm$ 4.41 \\\hline
$K_{s_2}$ & 38 & 1.9 & 38 $\pm$ 3.72 \\\hline
$K_{s_3}$ & 40 & 2.0 & 40 $\pm$ 3.92 \\\hline
$a$ & 1 & 0.06 & 1 $\pm$ 0.118\\\hline
$b$ & 0 & 100 & 0 $\pm$ 196\\\hline
$c$ & 0 & 900 & 0 $\pm$ 1760\\
\hline
\end{tabular}
\label{tab:PDFs}
\end{table}

\subsection{Ensemble-based Data assimilation algorithm} \label{subsect:algoEnKF}

Continuous time-series of gauged water levels and/or discharge recorded at discrete locations have traditionally been used for model calibration and validation as well as with DA algorithm for real-time constraint of hydraulic flood forecasting models (e.g., \cite{Madsen2005, Neal2007, Neal2009}). In the present work, gauged water levels are assimilated with an Ensemble Kalman Filter (EnKF) algorithm, for T2D Garonne model, in order to sequentially correct the friction and inflow discharge. The classical EnKF algorithm is favored as it allows to stochastically estimate the covariances between the model inputs/parameters and outputs, without formulating the tangent linear of the hydrodynamics model, under the assumption that the errors in the control vector are properly described by a gaussian probability density function.

\subsubsection{Description of the control vector}\label{subsect:EnKFctl}

The DA algorithm consists in a cycled stochastic EnKF, where the control vector ${\bf x}$ is composed of the friction coefficients (4 scalars $K_{s_i}, i\in [0,3]$) and parameters that modify the time-dependent upstream boundary condition (3 scalars $a, b, c$); $n$ denotes the size of the control vector. These 7 parameters are assumed to be constant over a DA cycle, yet their evolution in time is made possible by DA between cycles. The DA cycle $c$ covers a time window, denoted by $W_c = [t_{start}$;~$t_{end}]$ of length $T$ over which $N_{obs, c}$ in-situ observations are assimilated. The cycling of the DA algorithms consists in sliding the time window of a period $T_{shift}$ so that the cycles $c$ and $c+1$ may overlap. \\
It could be argued that the DA algorithm is more a smoother than a filter as it operates over a sliding time window. Yet, as the control vector is composed of model parameters (and not of the model state) that are assumed constant over the assimilation window, the smoothing resumes to a filtering. The EnKF algorithm relies on the propagation of $N_e$ members with perturbed values of ${\bf x}$, denoted by ${\bf x}^i$, i.e. the forecast values denoted by ${\bf x}^{f,i}_c$ (superscript index $f$ stands for \textquotedblleft forecast\textquotedblright), where $i \in [1, N_e]$ is the ensemble member counter. 

\subsubsection{Description of the EnKF forecast step}\label{subsect:EnKFfctstep}

The EnKF forecast step consists in the propagation in time, over $W_c$ of the control and model state vectors. The EnKF is applied to model parameters that, by definition, do not evolve in time over cycle $c$. 
The absence of propagative model for the control vector  implies that the forecast for the control vector at cycle $c$ should remain equal to the analysis at cycle $c-1$. Yet, in order to avoid ensemble collapse, artificial dispersion is introduced within the sampling with the addition of perturbations ${\boldsymbol{\theta}}$ to the mean of the analysis from the previous cycle $\overline{{\bf x}^{a}_{c-1}}$ (superscript index $a$ stands for \textquotedblleft analysis\textquotedblright). 
The forecast step thus reads: 

\begin{equation}
    {\bf x}^{f,i}_{c} = \left\{
    \begin{array}{ll}
    {\bf x}_{0} + {\boldsymbol{\theta}}^i_1 & \textnormal{ if } c=1\\[5pt]
    \overline{{\bf x}^{a}_{c-1}} + {\boldsymbol{\theta}}^i_c & \textnormal{ if } c>1\\
    \end{array}
    \right.
    \label{eq:ctlforecast}
\end{equation}
with \begin{equation}
    \overline{{\bf x}^a_{c-1}} = \frac{1}{N_e} \sum\limits_{\substack{i = 1}}^{N_e} {{\bf x}^{a,i}_{c-1}} \; \;  \in \mathbb{R}^{n}
    \label{eq:ens_mean_x}
    \end{equation}
and 
\begin{equation}
    {\boldsymbol{\theta}}^i_c \sim {\cal{N}}{({\bf 0}, {\sigma^i_c}^2)}
\end{equation}
where
\begin{equation}
    \sigma^i_c = \left\{
    \begin{array}{ll}
    \sigma_{\bf x} & \textnormal{ if } c=1\\[5pt]
    \lambda_1\sqrt{\dfrac{1}{N_e-1}\sum_{i=1}^{N_e}({\bf x}^{a,i}_{c-1} - \overline{{\bf x}^{a}_{c-1}})^2} + \lambda_2\sigma_{\bf x} & \textnormal{ if } c>1\\
    \end{array}
    \right.
\end{equation}

For the first cycle, the perturbed friction and upstream forcing coefficient values are drawn within the PDFs described in \autoref{tab:PDFs}. 
For the next cycles, the set of coefficients issued from the mean analysis at the previous cycle is further dispersed by additive perturbations ${\boldsymbol{\theta}}$ drawn from the Gaussian distribution with zero mean and a standard deviation obtained from the linear combination of the standard deviation of the analysis at the previous cycle and $\sigma_{\mathbf{x}}$ described in \autoref{tab:PDFs}. The two terms are weighted by the hyperparameters $\lambda_1$ and $\lambda_2$.
This technique is an advanced alternative to anomalies inflation for avoiding the well-known ensemble collapse, better suited for heterogeneous control of parameters. The combined update of the variance for the re-sampling of the parameters allows to preserve  part of the information from the background statistical description that may differ amongst the parameters and over time while also inheriting analyzed variance from the previous cycle. In the following implementation, $\lambda_1$ and $\lambda_2$ are respectively set to $0.3$ and $0.7$ after the analysis of the ensemble spread in the control space along the DA cycles.
The background hydraulic state, associated with each member of the ensemble of inputs, denoted by ${\bf s}^{f,i}_c$, results from the integration of the hydrodynamic model ${\cal{M}}_c$ : ${\mathbb{R}}^n \rightarrow {\mathbb{R}}^m$  from the control space to the model state (of dimension $m$) over $W_c$: 
\begin{equation}
{\mathbf{s}}^{f,i}_{c} = {\cal{M}}_{c}({\bf s}^{a,i}_{c-1},{\bf x}^{f,i}_{c})
\label{eq:stateforecast}
\end{equation}
The initial condition for ${\cal{M}}_{c}$ at $t_{start}$ is provided by a user-defined restart file for the first cycle. 
For the following cycles, it takes in a full restart ${\bf s}^{a,i}_{c-1}$ saved from the analysis run of the previous cycle ${\mathbf{s}}^{a,i}_{c-1} = {\cal{M}}_{c-1}({\bf s}^{a,i}_{c-2},{\bf x}^{a,i}_{c-1})$ at time $t_{start}+T_{shift}$. Note that in order to avoid inconsistencies between the state and the analysed set of parameters at $t_{start}$, a short spin-up integration is run on the 3 hours preceding $t_{start}$.
The control vector equivalent in the observation space for each member, denoted by ${\bf y}^{f,i}_c$, stems from: 
\begin{equation}
{\mathbf{y}}^{f,i}_{c} = {{\cal{H}}_c}({\mathbf{s}}^{f,i}_{c})
\label{eq:ctlequivobs}
\end{equation}
where ${{\cal{H}}_c}$ : ${\mathbb{R}}^m \rightarrow {\mathbb{R}}^{n_{obs}}$ is the observation operator from the model state space to the observation space (of dimension $n_{obs}$) that selects, extracts and eventually interpolates model outputs at times and locations of the observation vector $\mathbf{y}^o_c$ over $W_c$. It should be noted that, in the following, the observation operator may also include a bias removal step to take into account a systematic model error. Eq.~\eqref{eq:ctlequivobs} thus reads
\begin{equation}
{\mathbf{y}}^{f,i}_{c} = {{\cal{H}}_c}({\mathbf{s}}^{f,i}_{c}) - \mathbf{y}_{bias}
\label{eq:ctlequivobsbias}
\end{equation}
where $\mathbf{y}_{bias}$ is an a priori knowledge of the model-observation bias. The estimation of this bias is further described in \autoref{ssec:merits_FR1_FR2} for the studied flood event.

\subsubsection{Description of the EnKF analysis step}\label{subsect:EnKFanastep}

The EnKF analysis step stands in the update of the control and model state vectors. When applying a stochastic EnKF \cite{asch2016data}, the observation vector $\mathbf{y}^{o,i}$ is perturbed, and an ensemble of observations $\mathbf{y}^{o,i}_c$ ($i \in [1, N_e]$) is generated: 
\begin{align}
\mathbf{y}^{o,i}_c = \mathbf{y}^{o}_c + \boldsymbol{\epsilon}_c \; \; \; \textnormal{with} \; \; \boldsymbol{\epsilon}_c \sim {\cal N}({\bf 0},{\bf R}_c) 
\label{eq:pertobs}
\end{align}
where ${\bf R}_c= {\sigma_{obs}}^2 {\mathbf I}_{n_{obs}}$ is the observation error covariance matrix, here assumed to be diagonal, of standard deviation $\sigma_{obs}$ (and ${\mathbf I}_{n_{obs}}$ is the ${n_{obs} \times n_{obs}}$ identity matrix), as the observation errors are assumed to be uncorrelated, Gaussian and with a standard deviation a standard deviation proportional to the observations $\sigma_{obs,k} = \tau \mathbf{y}^{o}_c$. The innovation vector over $W_c$ is the difference between the perturbed observation vector $\mathbf{y}^{o,i}_c$ and the model equivalent $\mathbf{y}^{f,i}_c$ from Eq.~\eqref{eq:ctlequivobs} and Eq.~\eqref{eq:pertobs}. It is weighted by the Kalman gain matrix $\mathbf{K}_{c}$ and then added as a correction to the background control vector ${\bf x}^{f,i}_{c}$, so that the analysis control vector ${\bf x}^{a,i}_{c}$ is computed in Eq.~\eqref{eq:ctlana}.
\begin{equation}
{\bf x}_c^{a,i} = {\bf x}^{f,i}_{c} + \mathbf{K}_{c}\; ({\bf y}^{o,i}_{c} - {\bf y}^{f,i}_{c}).
\label{eq:ctlana}
\end{equation}
The Kalman gain reads:
\begin{equation}
\mathbf{K}_c = \mathbf{P}^{\bf{x},\bf{y}}_c {\left[ \mathbf{P}^{\bf{y},\bf{y}}_c + \mathbf{R}_{c} \right]}^{-1}
\label{eq:EnKF_ana_Klambda_gain_chap12}
\end{equation}
with $\mathbf{P}^{\bf{y},\bf{y}}_{c}$ the covariance matrix of the error in the background state equivalent in the observation space $\mathbf{y}^{f}_{c}$ and $\mathbf{P}^{\bf{x},\bf{y}}_{c}$  the covariance matrix between the error in the control vector and the error in $\mathbf{y}^{f}_{c}$, stochastically estimated within the ensemble:
\begin{align}
\mathbf{P}^{\bf{x},\bf{y}}_{c} & = \frac{1}{N_e} \mathbf{X}_c^T \mathbf{Y}_c  \;  \; \in \mathbb{R}^{n \times n_{obs}}\\
\mathbf{P}^{\bf{y},\bf{y}}_{c} & = \frac{1}{N_e} \mathbf{Y}_c^T \mathbf{Y}_c  \;  \; \in \mathbb{R}^{n_{obs} \times n_{obs}}
\label{eq:Pyy}
\end{align}
with:
\begin{align}
\mathbf{X}_c & = \left[ {\bf x}^{f,1}_c- \overline{{\bf x}^f_c}, \cdots, {\bf x}^{f,N_e}_{c} -\overline{{\bf x}^f_c}\right] \; \;   \in \mathbb{R}^{n \times N_e}\\
\mathbf{Y}_c & = \left[ {\bf y }^{f,1}_c- \overline{\bf y^f_c}, \cdots, {\bf y}^{f,N_e}_{c} - \overline{{\bf y}^f_c}\right] \; \;   \in \mathbb{R}^{n_{obs} \times N_e}
\label{eq:ens_anomaly_matrix}
\end{align}
and 
\begin{align}
\overline{{\bf x}^f_c} & = \frac{1}{N_e} \sum\limits_{\substack{i = 1}}^{N_e} {{\bf x}^{f,i}_c} \; \;  \in \mathbb{R}^{n}\\
\overline{{\bf y}^f_c} & = \frac{1}{N_e} \sum\limits_{\substack{i = 1}}^{N_e} {{\bf y}^{f,i}_c} \; \;  \in \mathbb{R}^{n_{obs}}.
\label{eq:ens_mean_y}
\end{align}

The analyzed hydrodynamic state, associated with each analyzed control vector ${\bf x}^{a,i}_c$ is denoted by ${\bf s}^{a,i}_c$. It results from the integration of the hydrodynamic model ${\cal{M}}_c$ with updated friction and upstream forcing $Q_{up}$ over $W_c$, starting from the same initial condition as each background simulation within the ensemble: 
\begin{equation}
{\mathbf{s}}^{a,i}_{c} = {\cal{M}}_{c}({\bf s}^{a,i}_{c-1},{\bf x}^{a,i}_{c}).
\label{eq:stateanalyzed}
\end{equation}

\subsubsection{Cycled forecast with DA}\label{subsect:Forecast}
The integration of the cycled ensemble forecast (of $N_e=24$ members in the following experiments) is initialized with the full restarts saved at the final time of the DA analysis runs (i.e. $t_{end}$) and uses the updated friction coefficients and the inflow corrections issued from the EnKF ${\bf x}^{a,i}_{c}$ for the following forecast duration (selected here as +24h).
The average state of this ensemble will be further used to assess the performance of DA in forecast mode.

\subsection{Flood extent mapping using SAR imagery data}\label{subsect:floodml}
 
Sentinel-1 is the first satellite series of the Copernicus program \cite{torres2012gmes}. This SAR system works at C-band, with a central frequency of 5.405 GHz. The Interferometric Wide (IW) mode used in this study offers a ground resolution of approximately 20 $\times$ 22 m; this product is then resampled, reprojected and distributed at  10 $\times$ 10 m for Ground Range Detected (GRD) products. In order to improve the revisit time, Sentinel-1 works as a constellation of two identical satellites Sentinel-1A launched on 2014-04-03 and Sentinel-1B on 2016-04-26, allowing a six-day revisit time. 
The Sentinel-1 GRD IW products are leveraged to produce binary water maps using Machine Learning algorithms developed by CNES and CLS in the frame of the FloodML project \cite{2020AGUFMIN041..09H,kettig}. 
FloodML aims at developing advanced Artificial Intelligence algorithms to improve the accuracy and decrease the computational time for flooded area retrieval from RS data. 

Random Forest (RF) is an ensemble learning method that is built on a multitude of decision tree classifiers on the sub-samples of a training dataset.
A decision is then made by counting all the tree votes and choosing the majority or the average of the responses, in order to improve the predictive accuracy and limit data over-fitting.
In the present work, a RF algorithm \cite{pal2005random,belgiu2016random} was trained over a dataset that gathers 223 Sentinel-1 images from 12 non-coastal Copernicus EMS Rapid mapping (EMSR) flood cases. The EMSR open dataset provides satellite EO data dedicated to the surveillance and the management of natural disasters, emergencies and humanitarian crises worldwide. In the present study, the training dataset consists in flood maps labeled with high-occurrence (greater than 90\%) water pixels selected from the Global Surface Water Occurrence reference \cite{pekel2016high}.

The use of both Sentinel-1's VV and VH polarizations was explored, in conjunction with the use of the local slope derived from MERIT Digital Elevation Model \cite{yamazaki2017high}. 
The VV and VH Sentinel-1 images are first calibrated and orthorectified, then inferred by the RF algorithm to produce the water binary map.
In this work, the RF is performed using cuML, an open-source GPU-accelerated machine learning library, which allows to generate a flood extent map within 3 minutes for an orthorectified and tiled Sentinel-1 image (size of 11 $\times$ 11 km). 
The number of classifiers is fixed to 100 with no tree depth limit. 
The trained RF algorithm has been validated on five test sites all over the world, and yield an average $F_2$-score of 86.86\%,
which is more accurate compared to several other tested Machine Learning methods such as Support Vector Machine and k-Nearest Neighbors.
A Deep Learning method, namely U-Net \cite{ronneberger2015u}, was also developed for this water detection task using Sentinel-1 and Sentinel-2 images, but it only achieves a similar level of accuracy compared to the RF algorithm, thus only RF is used in this paper owing to its computational efficiency.
In order to remove noises and fill the holes in the resulting detected flood surfaces, a majority filter (with the radius set equal to 3 pixels) has been applied on the resulting flood binary mask.

As aforementioned, this paper focuses on two major events that occurred over the Garonne Marmandaise in December 2019 and January-February 2021, where several Sentinel-1 images were acquired during the events. 
Three Sentinel-1 orbits (Ascending 30, Ascending 132 and Descending 8) provide observation and associated images over this catchment as detailed in \autoref{tab:events_info}.
 \begin{table}[h]
 \centering
  \caption{General information on the studied flood events.}
 \label{tab:events_info}
 \begin{tabular}{c|c|c|c}
    \hline
    Event & First date & Last date  & Nb of Sentinel-1 images \\ \hline
    2021 & 2021-01-16 & 2021-02-15  & 9 \\ \hline
    2019 & 2019-12-13 & 2019-12-29  & 8 \\ \hline
 \end{tabular}
 \end{table}

\begin{figure}[th]
     \centering
     \begin{subfigure}[b]{0.45\textwidth}
         \centering
         \includegraphics[width=\linewidth]{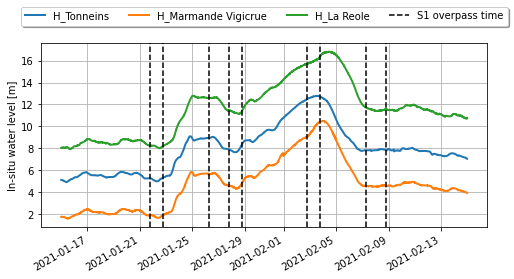}
         \caption{}
         \label{fig:all_obs_2021}
    \end{subfigure}
    \begin{subfigure}[b]{0.45\textwidth}
         \centering
         \includegraphics[width=\linewidth]{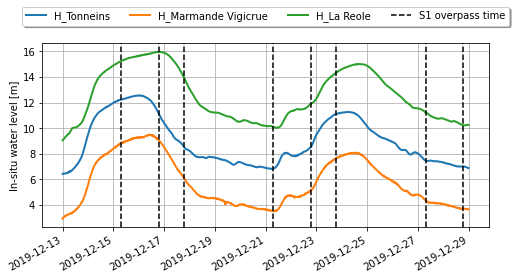}
         \caption{}
         \label{fig:all_obs_2019}
    \end{subfigure}
    
    \caption{Water level $H$ time-series for (a) 2021 January-February flood event, (b) 2019 December  flood event, at Tonneins (blue curve), Marmande (orange curve) and La Réole (green curve).  Sentinel-1 overpass times are indicated as vertical dashed lines.}
\end{figure}

\begin{figure*}[h!]
     \centering
     \begin{minipage}{0.45\textwidth}
     \begin{subfigure}[b]{\textwidth}
         \centering
         \includegraphics[width=\linewidth]{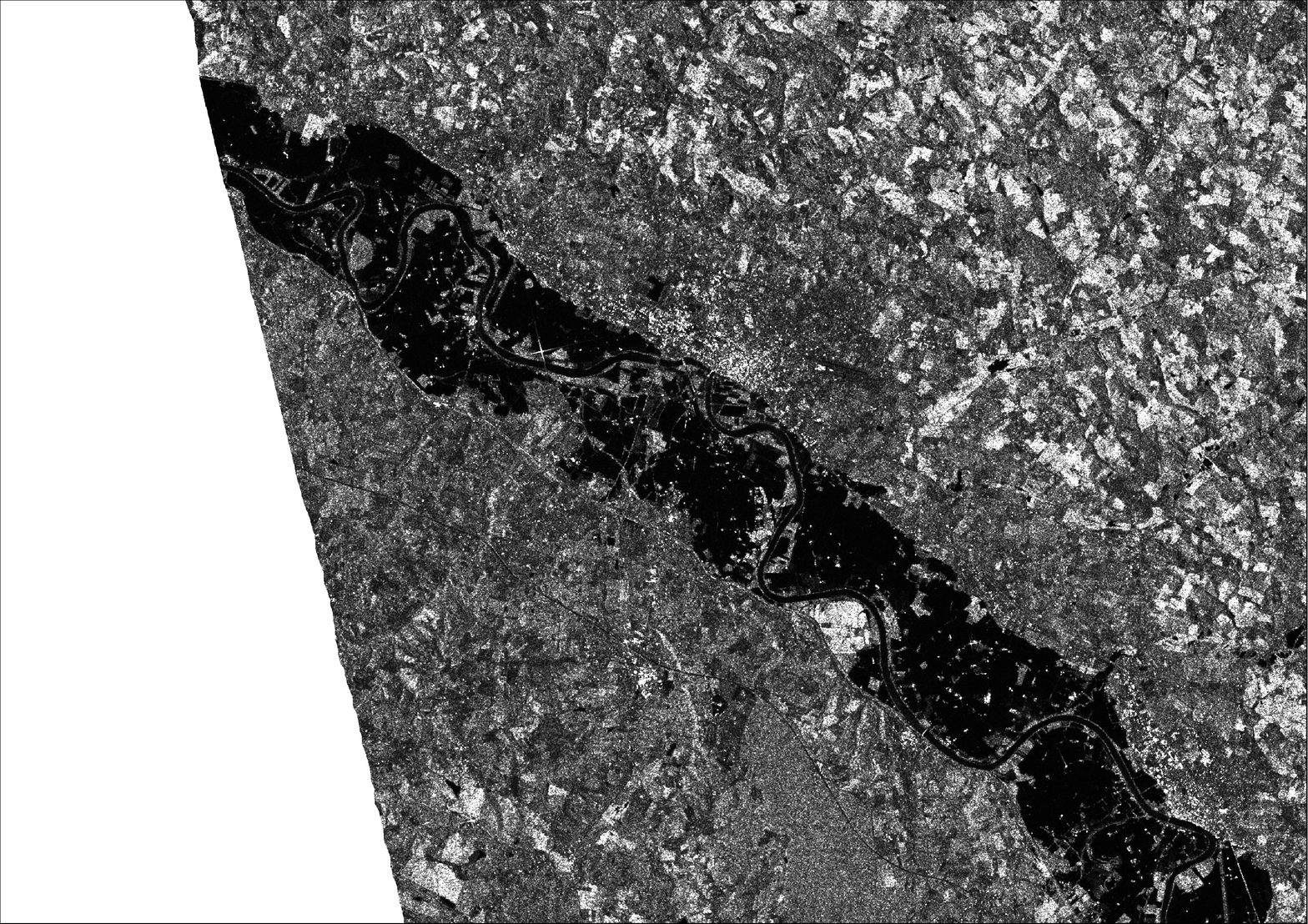}
         \caption{}
         \label{fig:FML+FR1+OSO16_2021_SAR}
     \end{subfigure}
     \begin{subfigure}[b]{\textwidth}
         \centering
         \includegraphics[width=\linewidth]{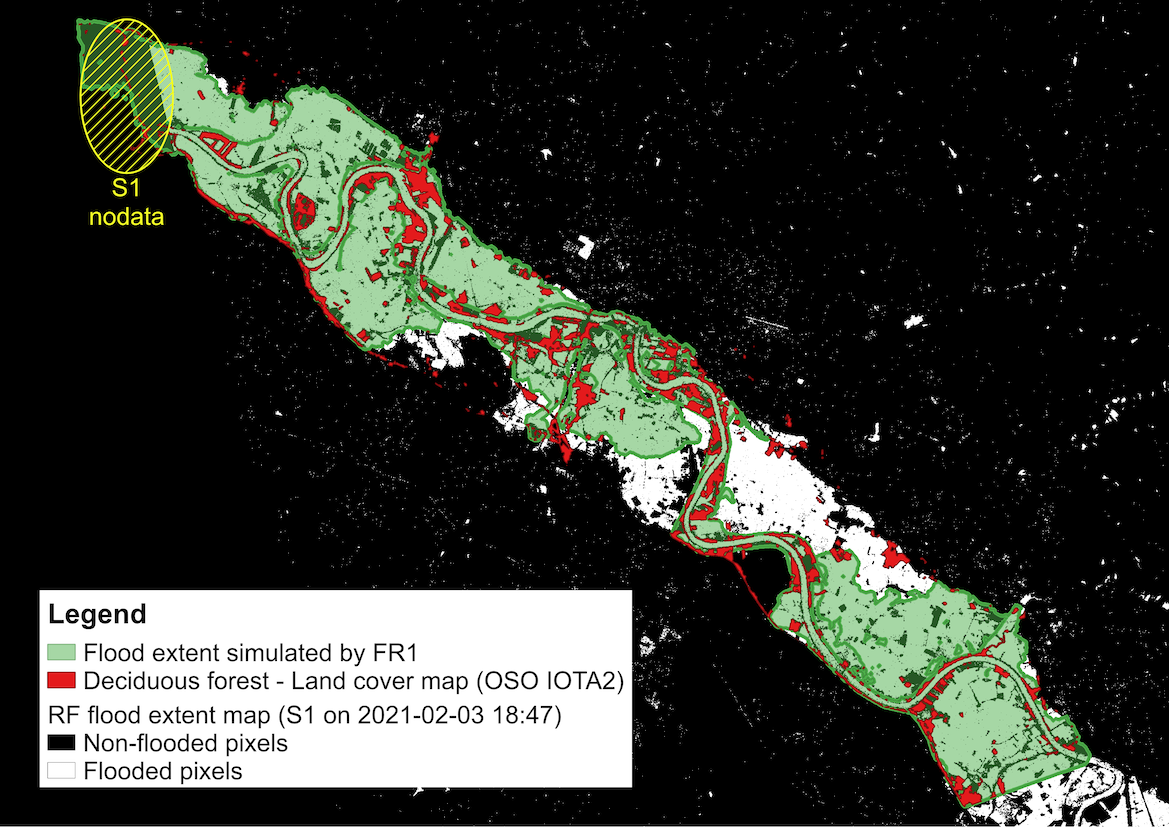}
         \caption{}
         \label{fig:FML+FR1+OSO16_2021_FloodML}
     \end{subfigure}
     \begin{subfigure}[b]{\textwidth}
         \centering
         \includegraphics[width=\textwidth]{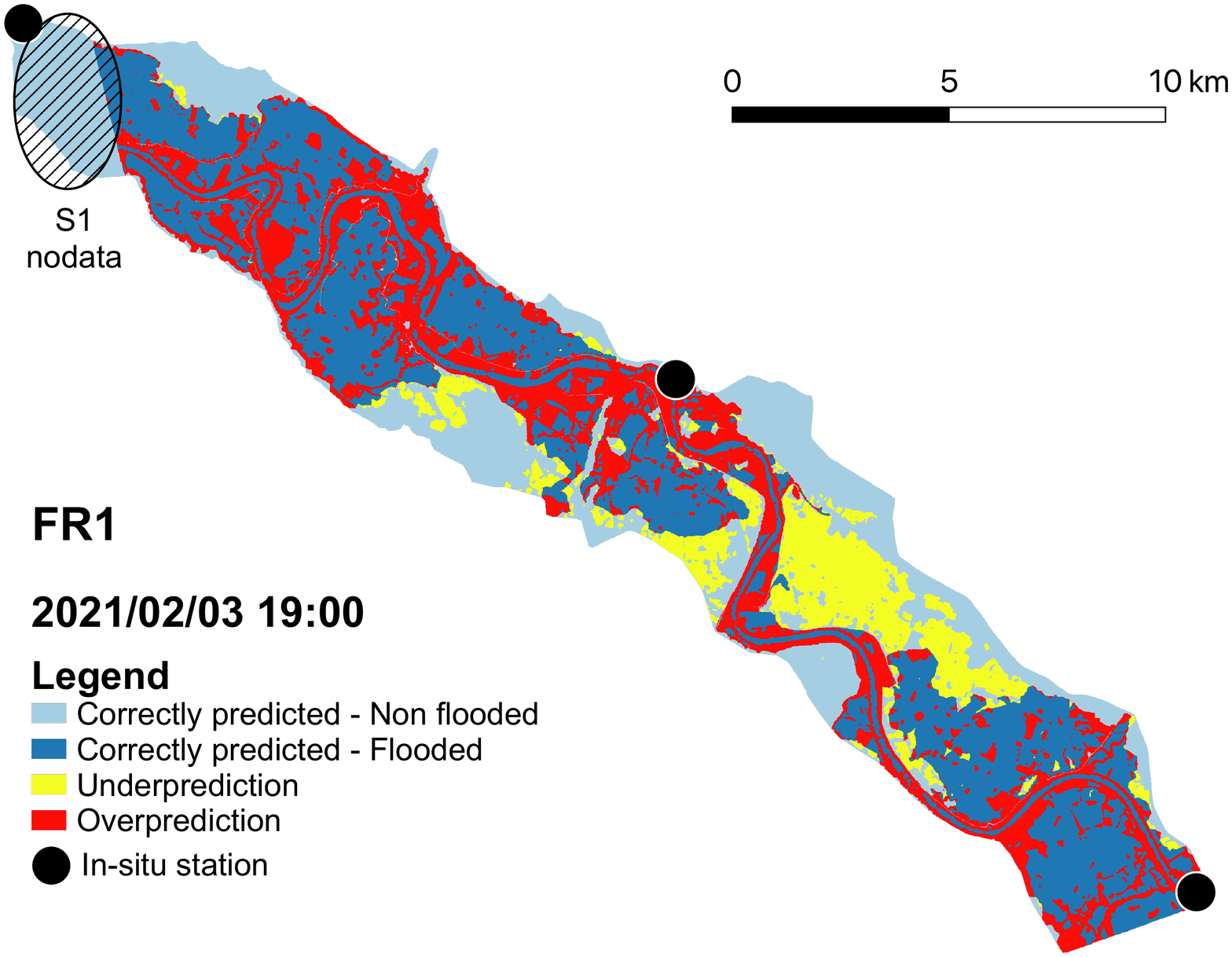}
         \caption{}
         \label{sfig:FreeRun_1710000_2021_pred}
     \end{subfigure}
     \end{minipage}
     \begin{minipage}{0.45\textwidth}
     \begin{subfigure}[b]{\textwidth}
         \centering
         \includegraphics[width=\linewidth]{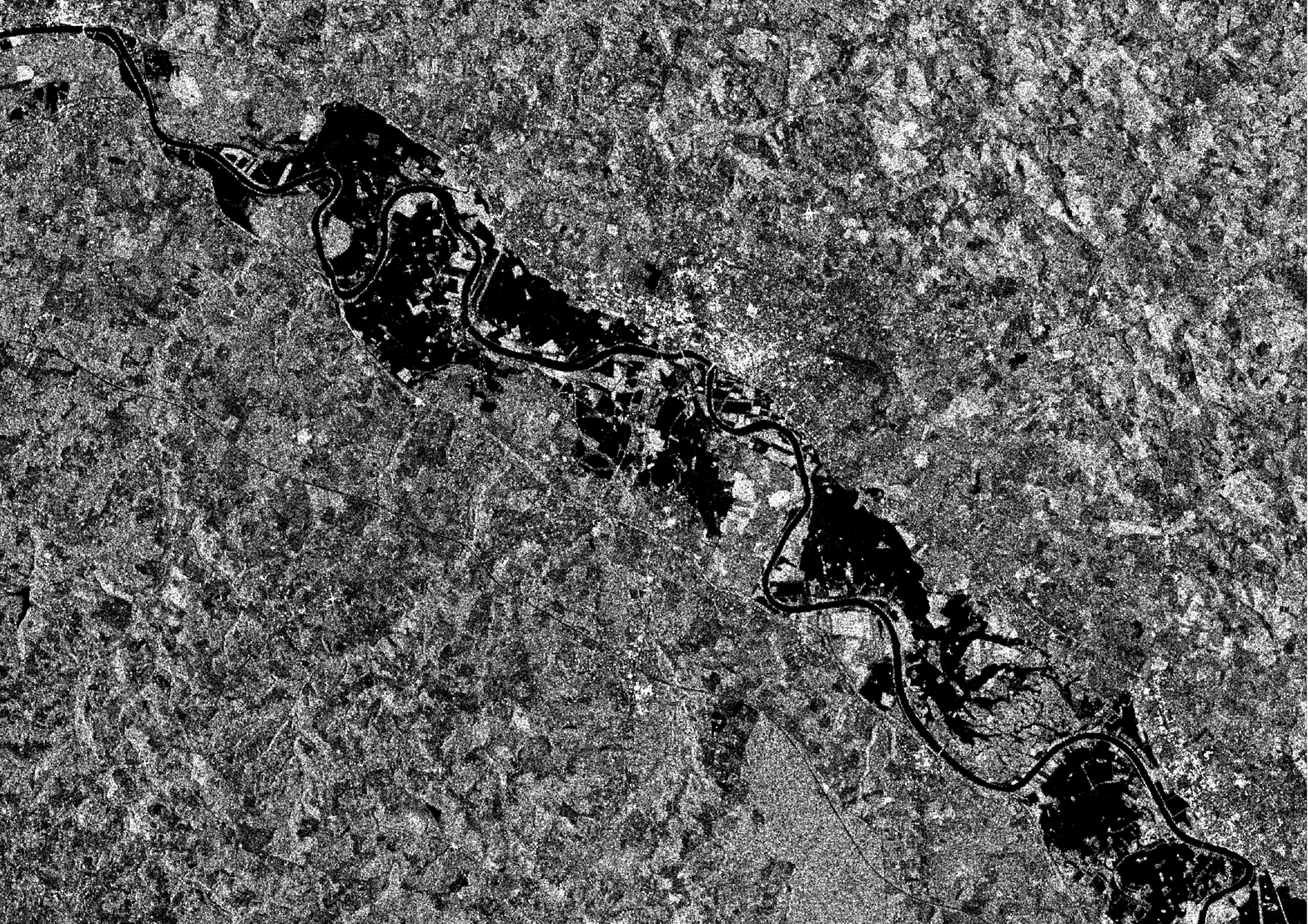}
         \caption{}
         \label{fig:FML+FR1+OSO16_2019_SAR}
     \end{subfigure}
     \begin{subfigure}[b]{\textwidth}
         \centering
         \includegraphics[width=\linewidth]{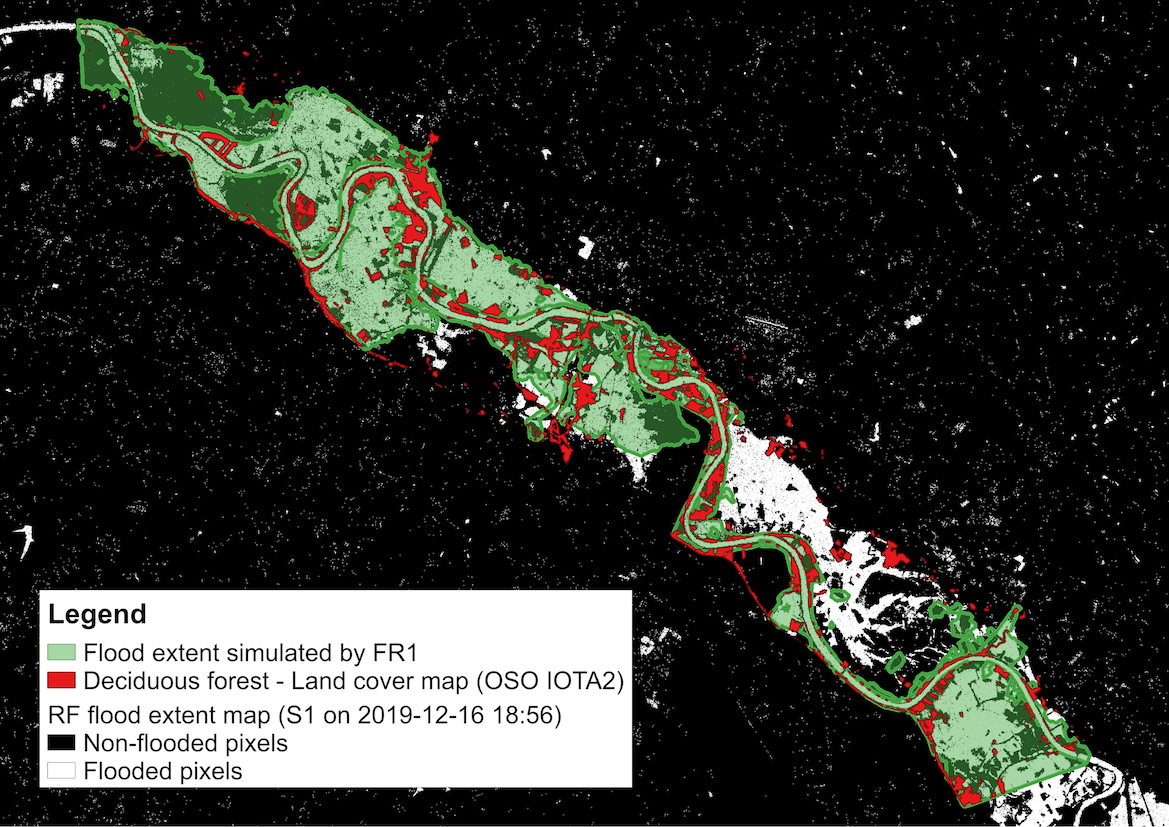}
         \caption{}
         \label{fig:FML+FR1+OSO16_2019_FloodML}
     \end{subfigure}
     \begin{subfigure}[b]{\textwidth}
         \centering
         \includegraphics[width=\textwidth]{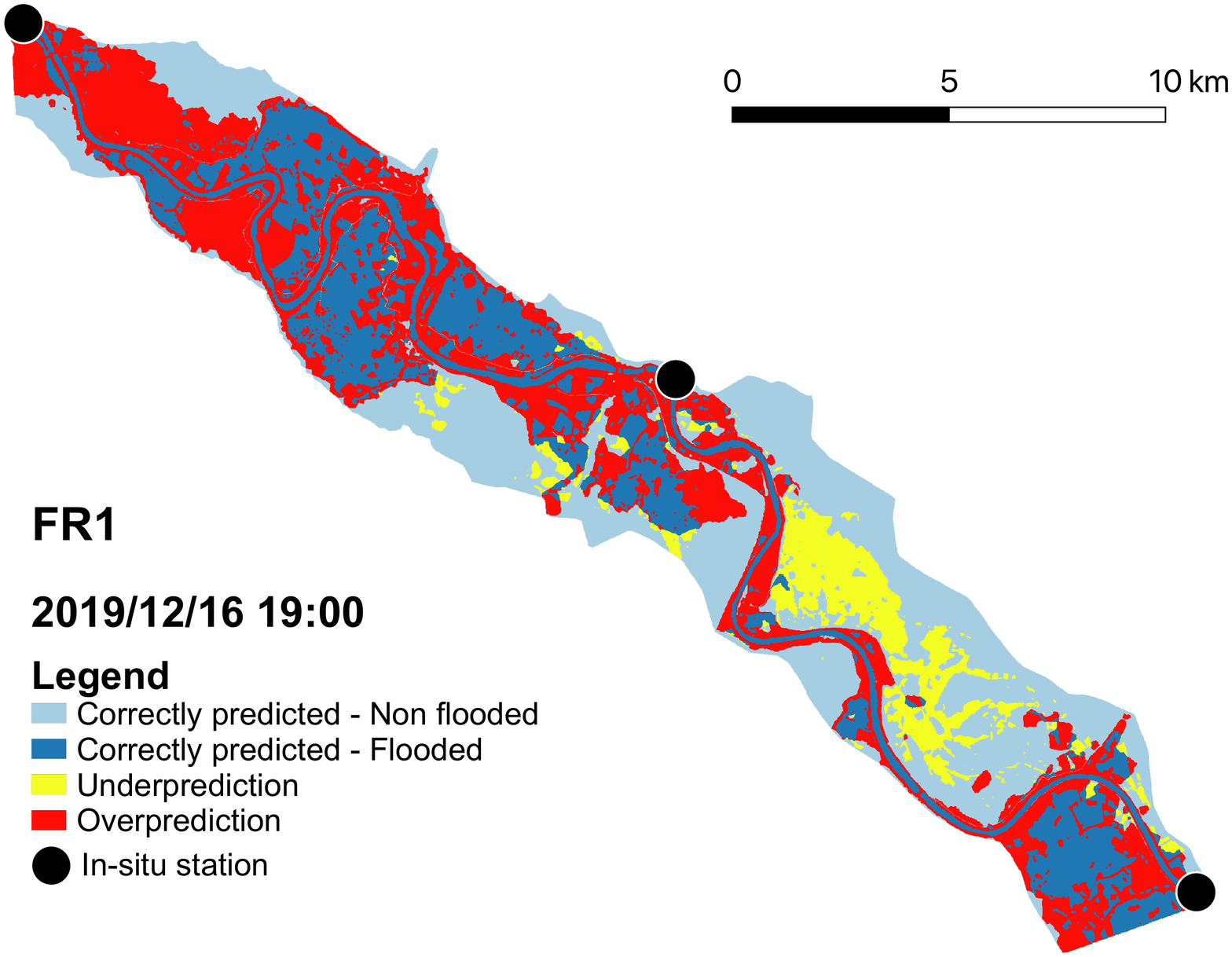}
         \caption{}
         \label{sfig:FreeRun_414000_2019_pred}
     \end{subfigure}
     \end{minipage}
     \caption{(a) Sentinel-1 SAR image for 2021-02-03 18:48 (respectively, (d) for 2019-12-16 18:56), (b)  FloodML and T2D flood extent maps of a free run for 2021-02-03 (respectively, (e) for 2019-12-16). Red regions represent the deciduous forest areas to be excluded from the comparison between flood extents. (c) Contingency map representing the flood prediction by free run for 2021-02-03 (respectively, (f) for  2019-12-16). The correctly predicted flooded areas are represented in dark blue, correctly predicted non-flooded areas in light blue, under predicted areas in yellow, over predicted areas in red.}
     \label{fig:FloodML}
\end{figure*}

\autoref{fig:all_obs_2021} (respectively \autoref{fig:all_obs_2019}) depicts the in-situ water level time-series observed for the 2021 (respectively 2019) event at Vigicrue observing stations: Tonneins (blue curve), Marmande (orange curve) and La Réole (green curve), with the Sentinel-1 overpass times indicated as vertical black dashed lines. The 2021 event is composed of a single peak and observed by 9 Sentinel-1 images. 
The flood peak of the 2021 event is covered by the ascending orbit 30 on 2021-02-02 18:55 and by the ascending orbit  132 the next day on 2021-02-03 18:48.
The 2019 event is composed of two peaks and observed by 8 Sentinel-1 images, where the first peak was observed by the image acquired on 2019-12-16 18:56. 
The Sentinel-1 images acquired close to the flood peaks on 2021-02-03 and 2019-12-16 are shown in \autoref{fig:FML+FR1+OSO16_2021_SAR} and \autoref{fig:FML+FR1+OSO16_2019_SAR}. 
It should be noted that for the Sentinel-1 images from the ascending orbit 132, a small part of the downstream area (including La Réole) is a no-data area as it is out of range from the acquisition and indicated by the hashed ellipse in \autoref{fig:FML+FR1+OSO16_2021_FloodML}.
The flood extent maps derived from FloodML on 2021-02-03 and 2019-12-16 are shown, respectively, in \autoref{fig:FML+FR1+OSO16_2021_FloodML} and \autoref{fig:FML+FR1+OSO16_2019_FloodML}. These flood extents are presented by binary masks, in which the white pixels (value 1) and black pixels (value 0) respectively indicate the flooded and non-flooded areas.  

As mentioned in \autoref{ssec:rsdata}, the flood detection using SAR images can be compromised by the presence of vegetated and urban regions. 
Over the Garonne Marmandaise catchment, this reduces the precision of the Random Forest inference results in regions where the vegetation is dense; here deciduous forest regions are indicated in red (\autoref{fig:FloodML}) with respect to the 2019 land cover map on French territory produced by the IOTA2 processing chain \cite{rs12030513}. 
As a matter of fact, these vegetated regions occlude the possible underflowing water when flood occurs---as only the tree trunks are submerged ---and present typical backscatter of deciduous forests on SAR images. A method dealing the detection of flood in different vegetation areas on SAR images can be found in \cite{grimaldi2020}. Hence, it is necessary to exclude these regions when comparing the simulated and observed flood extents.
These regions take up 8.6\% of the whole Garonne Marmandaise catchment total area.

\subsection{Flood modelling assessment metrics}\label{subsect:metrics}
In order to evaluate the flood modelling results, two assessment methods are carried out. They involve using in-situ data and Sentinel-1 derived flood extent maps for, respectively, assessing the simulated water level time-series and assessing the simulated flood extents. It should be recalled that only in-situ water levels are assimilated in the present work, a significant improvement with respect to these data is thus expected from DA analysis. Yet, Sentinel-1 derived flood extent maps are not assimilated, an improvement in metrics with respect to these independent data thus constitutes a more difficult goal to achieve for the DA strategy.

\subsubsection{1D metrics for water level time-series assessment}\label{sect:metrics}
The quality of the simulated water level $H^m$ is assessed with respect to in-situ observations $H^o$ in terms of mean and maximum computing the root-mean-square error ($\mathrm{RMSE}$), the maximum absolute error ($\mathrm{MaAE}$) and the  Nash–Sutcliffe model efficiency coefficient ($\mathrm{NSE}$) over the event time-series:
\begin{itemize}
    \item $\mathrm{RMSE}$ is computed between the simulated and the observed water level time-series:
        \begin{equation}
         \mathrm{RMSE} = \sqrt{\dfrac{1}{n_{obs}} \sum_{i=1}^{n_{obs}}(H_i^m-H_i^o)^2}   
        \end{equation}
     \item $\mathrm{MaAE}$ measures the maximum absolute difference between the water level time-series:
        \begin{equation}
         \mathrm{MaAE} = \max\limits_{i \in [1, n_{obs}]} ~\lvert H_i^m-H_i^o \rvert 
        \end{equation}   
    \item $\mathrm{NSE}$ reflects on the ratio of the error variance of the simulated time-series divided by the variance of the observed time-series:
        \begin{equation}
        \mathrm{NSE}=1-\dfrac{\sum_{i=1}^{n_{obs}}(H_i^m-H_i^o)^2}{\sum_{i=1}^{n_{obs}}(H_i^o-\bar{H^o})^2}
        \end{equation}
\end{itemize}
where $\bar{H^o}$ denotes the time-averaged observed water level over the event. For a perfect model, the estimation error variance computed with respect to observation is equal to 0 (i.e. $\sum_{i=1}^{n_{obs}}(H_i^m-H_i^o)^2 = 0$), thus the resulting $\mathrm{NSE}$ is equal to 1. 
Resulting $\mathrm{NSE}$ values nearer to 1 suggest a model with more predictive capacity.
A model that produces an estimation error variance equal to the variance of the observed time-series results in a $\mathrm{NSE}$ equal to 0.
Furthermore, when the estimation error variance computed with respect to observation is larger than the variance of the observations, the $\mathrm{NSE}$ becomes negative.
In other words, an efficiency less than zero ($\mathrm{NSE} < 0$) occurs when the observed mean is a better predictor than the model. 

\subsubsection{2D metrics for flood extent assessment}
The simulated flood extent maps are generated from the T2D simulated water level 2D field, by applying a threshold of $0.05$ m below which the pixel is considered as dry and above which it is considered as wet. The T2D water level output field is projected onto the regular grid of the Sentinel-1 image and FloodML inference map (ground sampling distance: 10 $\times$ 10 m) so that the two flood extent maps are comparable. 
The metrics to compare the simulated and the observed flood extent are: Critical Success Index ($\mathrm{CSI}$), $F_1$-score, and Cohen's kappa index ($\kappa$). 
$\mathrm{CSI}$ and $F_1$-score consider the FloodML flood extent maps as the reference observed flood maps (ground truth) based on which the T2D simulated flood extent maps are evaluated, whereas the objective of $\kappa$ index is to measure the agreement between the two flood extent estimators. The formulation of these indices relies on the count of pixels following one of four outcomes, color-coded in \autoref{sfig:FreeRun_1710000_2021_pred} and \autoref{sfig:FreeRun_414000_2019_pred}: True Positives ($TP$, blue pixels) is the number of pixels correctly predicted as flooded, False Positives ($FP$, red pixels) or \textit{over-prediction} is the number of non-flooded pixels incorrectly predicted as flooded, True Negatives ($TN$, light blue pixels) is the number of pixels correctly identified as non-flooded, and False Negatives ($FN$, yellow pixels) or \textit{under-prediction} is the number of missed flooded pixels. 
Based on these counts, the $\mathrm{CSI}$, $F_1$-score, and $\kappa$ indices are computed as follows:

\begin{equation}
\mathrm{CSI}=\dfrac{TP}{TP+FP+FN}
\end{equation}

\begin{equation}
F_\beta = (1+\beta^2) \times \dfrac{\mathrm{precision} \times \mathrm{recall}}{\beta^2 \times \mathrm{precision} + \mathrm{recall}}
\end{equation}
where $\mathrm{precision} = \sfrac{TP}{(TP+FP)}$ and $ \mathrm{recall} = \sfrac{TP}{(TP+FN)}$. The $F_1$-score (with $\beta = 1$) is selected as the balanced mean between the $\mathrm{precision}$ and $\mathrm{recall}$.

\begin{equation}
\kappa=\dfrac{p_o-p_e}{1-p_e}
\end{equation}
where $p_o$ is the observed proportionate agreement and $p_e$ is the probability of a random agreement, defined as follows:
\begin{equation}\nonumber
\begin{array}{rl}
p_o = &\dfrac{TP+TN}{TP+FP+FN+TN} \\[10pt]
p_e = &\dfrac{TP+FN}{TP+FP+FN+TN}\times\dfrac{TP+FP}{TP+FP+FN+TN}.
\end{array}
\end{equation}
These metrics range from 0\% when there is no common area (i.e. no agreement) between the simulation (T2D) and observation (FloodML), and reach their highest value of 100\% when the prediction provide a perfect fit to the observed flood extents. It should be noted that the magnitude and the size of the flood (and consequently the number of pixels used for the computation) were shown by \cite{stephens2014problems} 
to have a significant influence on these indices; thus limiting their use for different event and different catchment comparison. Here, these indices are mostly used to compare different numerical experiments on a single catchment and on the same event.

 \section{Experimental Results}\label{sect:Results}
Four experiments were carried out with the T2D Garonne Marmandaise model, for each flood event. They either consist in a Free Run (FR) meaning that no data are assimilated, or in a Data Assimilation (DA) run. They also differ depending on whether a model-observation bias is taken into account in the observation operator, on the ensemble size and on the observation error variance. The experiments are described in \autoref{tab:runs}.
 
 \begin{table}[h]
     \centering
     \begin{tabular}{c|c|c|c|c}
        \hline
         Exp.  & Bias & Data  & Nb of & \\ 
         name & correction & Assimilation &  members $N_e$ & $\tau $ (\%)\\ \hline
         FR1 & No & No & 1 & - \\ \hline
         FR2 & Yes & No & 1 & - \\ \hline
         DA1 & No & Yes & 24 & 15 \\ \hline
         DA2 & Yes & Yes & 24 & 15 \\ \hline
     \end{tabular}
     \caption{Summary of the experiments realized for the 2021 and 2019 flood events.}
     \label{tab:runs}
 \end{table}
 
\begin{figure*}
     \centering
     \begin{minipage}{0.45\textwidth}
     \centering
     \begin{subfigure}[b]{\textwidth}
         \centering
         \includegraphics[width=\textwidth]{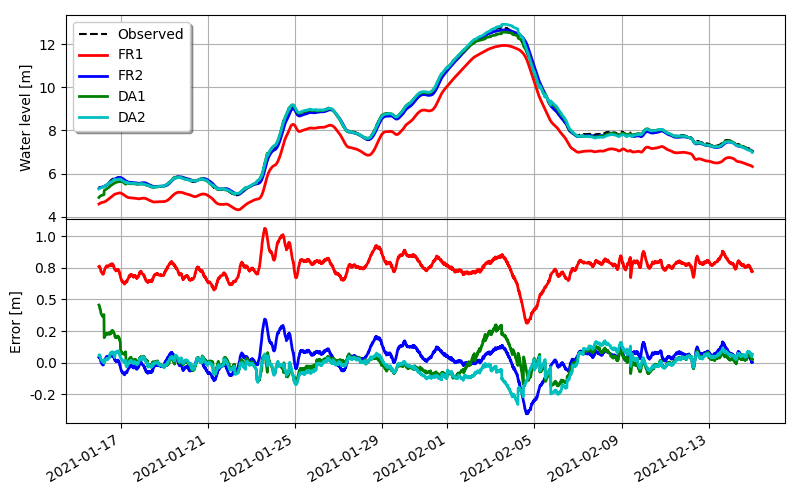}
         \caption{Tonneins}
         \label{fig:VigicrueFR1FR2DA2_Tonneins_2021}
     \end{subfigure}
     
     \begin{subfigure}[b]{\textwidth}
         \centering
         \includegraphics[width=\textwidth]{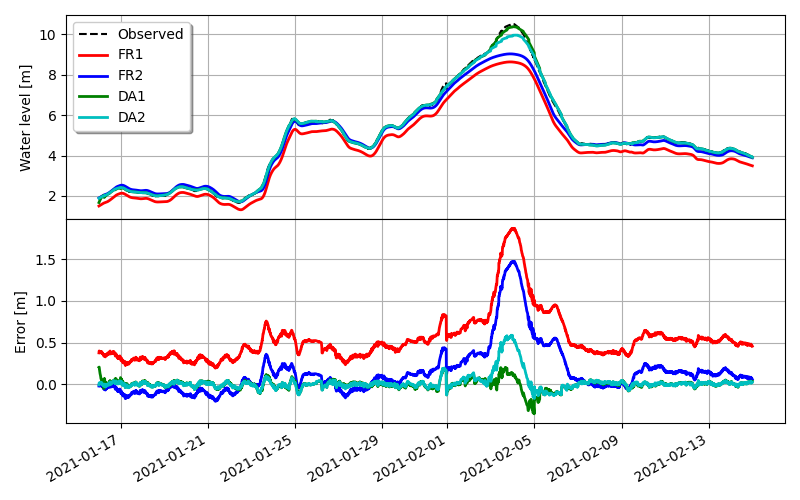}
         \caption{Marmande}
         \label{fig:VigicrueFR1FR2DA2_Marmande_2021}
     \end{subfigure}
     
     \begin{subfigure}[b]{\textwidth}
         \centering
         \includegraphics[width=\textwidth]{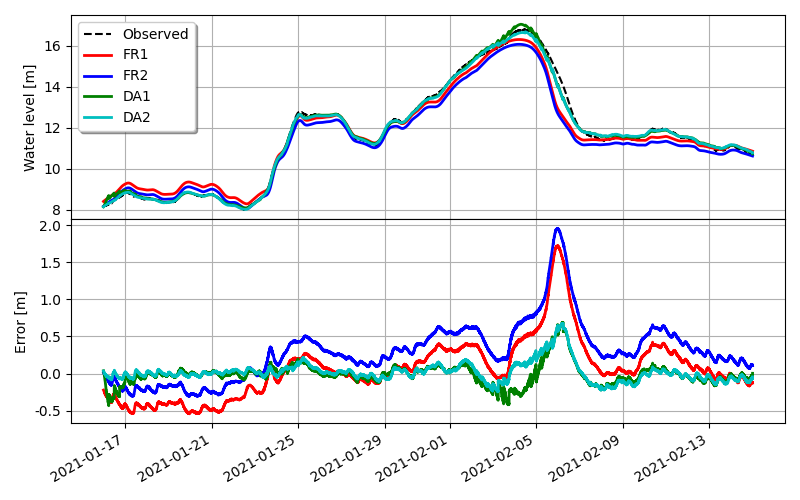}
         \caption{La Réole}
         \label{fig:VigicrueFR1FR2DA2_LaReole_2021}
     \end{subfigure}
     \end{minipage}
     \hfill
     \begin{minipage}{0.45\textwidth}
     \centering
     \begin{subfigure}[b]{\textwidth}
         \centering
         \includegraphics[width=\textwidth]{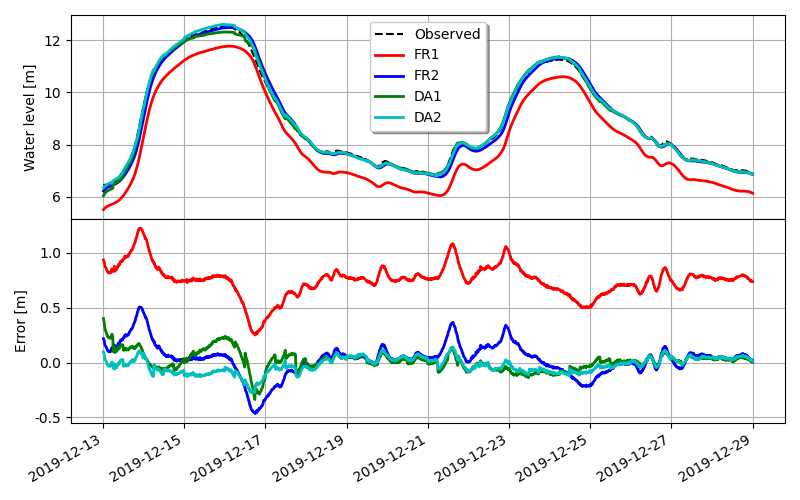}
         \caption{Tonneins}
         \label{fig:VigicrueFR1FR2DA2_Tonneins_2019}
     \end{subfigure}
     
     \begin{subfigure}[b]{\textwidth}
         \centering
         \includegraphics[width=\textwidth]{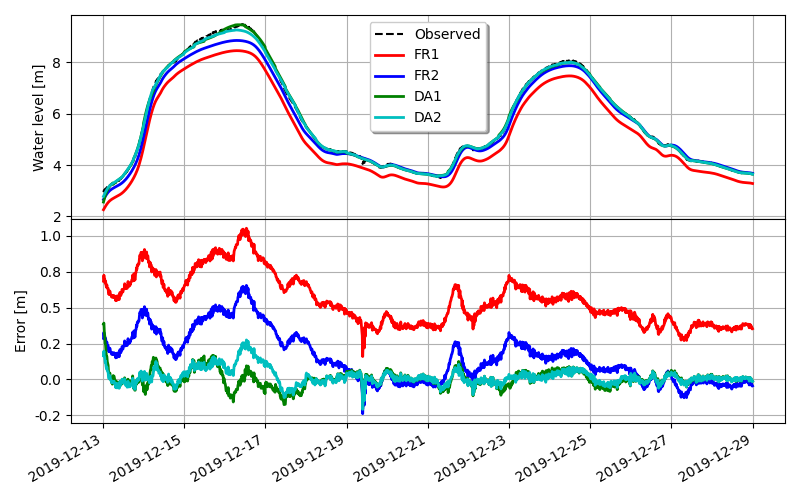}
         \caption{Marmande}
         \label{fig:VigicrueFR1FR2DA2_Marmande_2019}
     \end{subfigure}
     
     \begin{subfigure}[b]{\textwidth}
         \centering
         \includegraphics[width=\textwidth]{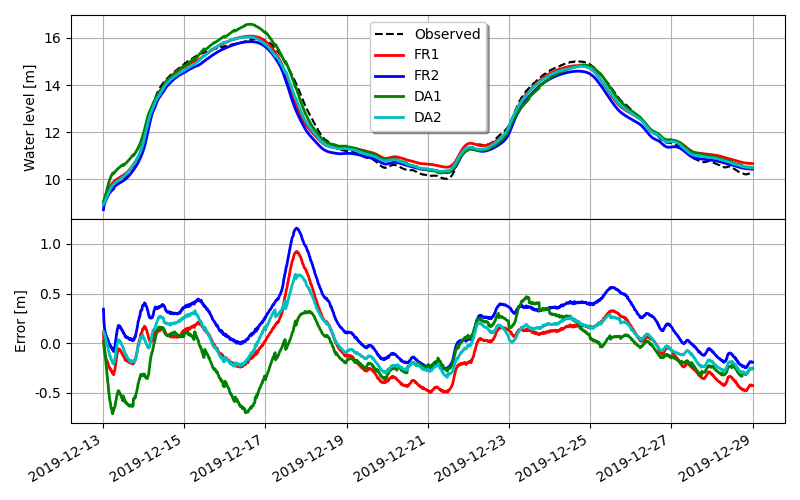}
         \caption{La Réole}
         \label{fig:VigicrueFR1FR2DA2_LaReole_2019}
     \end{subfigure}
     \end{minipage}
     \caption{Water level $H$ (upper plots) for Free run simulation (FR1 in red, FR2 in blue, DA1 in green and FR2 in cyan) and their respective error with respect to the observed values (lower plots) at Vigicrue in-situ observing stations (a) Tonneins, (b) Marmande, and (c) La Réole for January-February 2021 event on the left column (respectively, (d)-(f) for December 2019 event on the right column).}
     \label{fig:VigicrueFR1FR2DA2}
\end{figure*}

\autoref{fig:VigicrueFR1FR2DA2} displays the water levels at the 3 Vigicrue observing stations, Tonneins, Marmande, and La Réole, including the in-situ observed water levels (black dashed curve) and the simulated ones from the four different experiments (solid curves). 
FR1, FR2, DA1, and DA2 results are depicted respectively by red, blue, green, and cyan curves.
The errors between the observed and respective simulated water levels are also shown in the lower plot panels with the same color code.
Based on these water level time-series, the 1D assessment metrics (presented in \autoref{subsect:metrics}) are computed and summarized in \autoref{tab:stats_2021} and \autoref{tab:stats_2019}, respectively for 2021 and 2019 flood events.
In the following subsections, the merits of using RS data for model assessment are presented; with a focus on FR experiments in \autoref{ssec:merits_FR1_FR2}, on DA experiments in reanalysis mode in \autoref{ssec:merits_DA1_DA2}, and on DA experiments in forecast mode in \autoref{sec:forecast}.

 \subsection{Merits of RS data for free run model assessment - Experiments FR1 and FR2}
\label{ssec:merits_FR1_FR2}
\subsubsection{Water level at Vigicrue observing stations - Experiments FR1 and FR2}
 
The comparison of water level resulting from the T2D Free Run simulation FR1 with in-situ observations reveals that a systematic bias $\mathbf{y}_{bias}$ should be taken into account, as FR1 (red solid curve) significantly underestimates the observation (black dashed curve) over the entire flood events as shown in \autoref{fig:VigicrueFR1FR2DA2_Tonneins_2021} at Tonneins, \autoref{fig:VigicrueFR1FR2DA2_Marmande_2021} at Marmande and \autoref{fig:VigicrueFR1FR2DA2_LaReole_2021} at La Réole. For each sub-figure, the top panel represents the water level and the bottom panel represents the difference between the observation and the simulation. The bias was thus estimated during the first 24 hours on 2021-01-15, when the flow is  quasi-stationary and non-overflowing.
It is estimated at each observing station: $\mathbf{y}_{bias, Tonneins} = 72$ cm, $\mathbf{y}_{bias, Marmande} = 40$ cm, and $\mathbf{y}_{bias, La Reole} = -23$ cm. Water level results from the FR simulation FR2 that takes into account this bias in the comparison to the observation (Eq.~\eqref{eq:ctlequivobsbias}), are plotted with a blue solid curve in \autoref{fig:VigicrueFR1FR2DA2}. This simulation provides significantly improved results with respect to in-situ observations. The model-observation bias estimated from 2021 was used to carry out Free Run simulations over 2019 flood event and similar conclusions are drawn as illustrated in \autoref{fig:VigicrueFR1FR2DA2_Tonneins_2019} at Tonneins, \autoref{fig:VigicrueFR1FR2DA2_Marmande_2019} at Marmande and \autoref{fig:VigicrueFR1FR2DA2_LaReole_2019} at La Réole.  It is worth-noting that while the 2021-diagnosed bias is coherent with the model-observation misfit in 2019 at Tonneins and Marmande, there is a lesser agreement at La Réole. 
 
 \subsubsection{Rating curves for free runs - Experiments FR1 and FR2}
 
\begin{figure*}[t]
 \centering
    \begin{subfigure}[b]{0.44\textwidth}
    \centering
    \includegraphics[width=\linewidth]{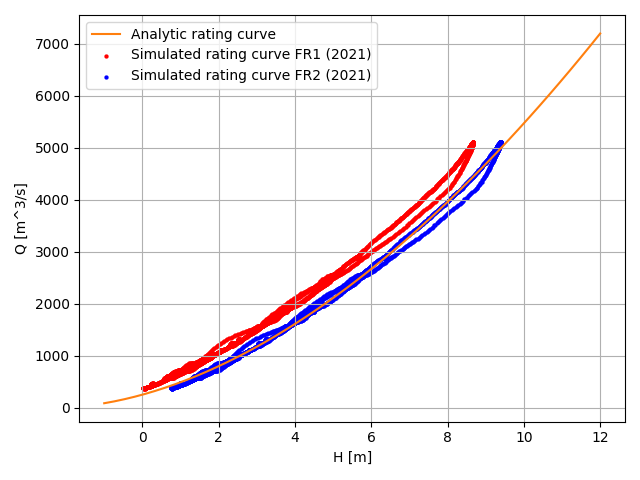}
    \caption{2021 flood event}
    \label{fig:CTTonneins_2021}
    \end{subfigure}
    \hfill
    \begin{subfigure}[b]{0.44\textwidth}
    \centering
    \includegraphics[width=\linewidth]{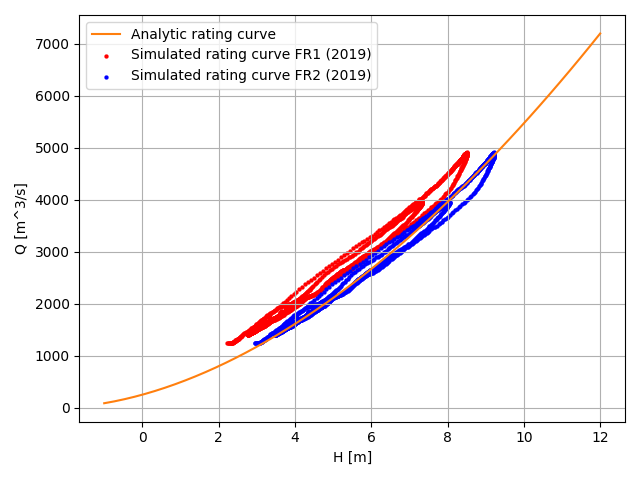}
    \caption{2019 flood event}
    \label{fig:CTTonneins_2019}
    \end{subfigure}
    \caption{Rating curves at Tonneins from analytical fit to observation (orange curve), and from the T2D simulations for FR1 (red curve) and for FR2 (blue curve) experiments. (a) 2021 flood event, (b) 2019 flood event.}
\end{figure*}

 The rating curve resulting from FR1 for the 2021 (respectively 2019) flooding event shown with a red curve in  \autoref{fig:CTTonneins_2021} (respectively \autoref{fig:CTTonneins_2019}) at Tonneins is not in agreement with the observation-based rating curve (orange curve) that was used to convert the observed water levels at Tonneins into discharge inflow to the model (previously mentioned in \autoref{subsect:T2D}). For both events, the bias correction applied to FR2 (blue curves) associates a given inflow to a larger water level , and consequently improves the coherence with the observed rating curve. The bias correction thus allows to improve the hydraulic variables at Tonneins where the upstream boundary condition is prescribed.
 
 \begin{table*}[t] 
\centering
\caption{1D scores with respect to in-situ data measured at Vigicrue observing stations.}
\begin{subtable}[t]{\textwidth}
\centering
\caption{1D scores - 2021 flood event}
\label{tab:stats_2021}
\begin{tabular}{c|ccc|ccc|ccc}
\hline
& \multicolumn{3}{c|}{{Root-mean-square error [m]}} & \multicolumn{3}{c|}{{Max absolute error [m]}} & \multicolumn{3}{c}{{Nash–Sutcliffe model efficiency}} \\
\hline
& Tonneins & Marmande & La Réole & Tonneins & Marmande & La Réole & Tonneins & Marmande & La Réole \\
\hline
{FR1} & 0.756 & 0.625 & 0.409 & 1.062 & 1.870 & 1.721 & 0.866 & 0.923 & 0.971 \\
\hline
{FR2} & 0.102 & 0.338 & 0.505 & 0.404 & 1.472 & 1.956 & 0.998 & 0.978 & 0.956\\
\hline
{DA1} & 0.090 & 0.059 & 0.148 & 0.456 & 0.352 & 0.690 & 0.869 & 0.948 & 0.868 \\
\hline
{DA2} & 0.084 & 0.104 & 0.138 & 0.330 & 0.590 & 0.676 & 0.892 & 0.917 & 0.889 \\
\hline
\end{tabular}
\end{subtable}\\
\vspace{0.25cm}
\begin{subtable}[t]{\textwidth}
\centering
\caption{1D scores - 2019 flood event}
\label{tab:stats_2019}
\begin{tabular}{c|ccc|ccc|ccc}
\hline
& \multicolumn{3}{c|}{{Root-mean-square error [m]}} & \multicolumn{3}{c|}{{Max absolute error [m]}} & \multicolumn{3}{c}{{Nash–Sutcliffe model efficiency}} \\
\hline
& Tonneins & Marmande & La Réole & Tonneins & Marmande & La Réole & Tonneins & Marmande & La Réole \\
\hline
{FR1} & 0.764 & 0.579 & 0.286 & 1.224 & 1.051 & 0.923 & 0.827 & 0.908 & 0.980 \\
\hline
{FR2} & 0.150 & 0.233 & 0.349 & 0.507 & 0.654 & 1.158 & 0.993 & 0.985 & 0.969\\
\hline
{DA1} & 0.089 & 0.058 & 0.279 & 0.401 & 0.392 & 0.710 & 0.802 & 0.911 & 0.446 \\
\hline
{DA2} & 0.074 & 0.060 & 0.224 & 0.273 & 0.275 & 0.695 & 0.833 & 0.918 & 0.632 \\
\hline
\end{tabular}
\end{subtable}
\end{table*}

\subsubsection{$\mathrm{RMSE}$, maximum absolute error and $\mathrm{NSE}$ at Vigicrue observing stations - Experiments FR1 and FR2}

 The dynamic of the river bed is assessed with respect to in-situ data at Tonneins, Marmande and La Réole. The $\mathrm{RMSE}$, $\mathrm{MaAE}$ and $\mathrm{NSE}$ measurements for 2021 event are given in \autoref{tab:stats_2021}. 
 When the bias is not taken into account in the comparison between free run results (FR1) and the in-situ observations, the $\mathrm{RMSE}$ over the 2021 event are 75.6 cm, 62.5 cm and 40.9 cm at Tonneins, Marmande and La Réole, respectively. 
 When the bias is removed, in FR2, the $\mathrm{RMSE}$ is reduced to 10.2 cm and 33.8 cm at Tonneins and Marmande. Yet, the estimation of the bias during the beginning of the first day of the event does not allow for an improvement over the entire event at La Réole as the nature of the error between the free run and the observation is not stationary at this location. Thus, the FR2 results are not improved with respect to FR1 at La Réole and the RMSE reaches  50.5 cm. 
 Significant reduction of the $\mathrm{MaAE}$ and improvement of the $\mathrm{NSE}$ measurements can also be remarked at Tonneins and La Réole.
While the model-observation bias correction in FR2 allows some improvements in terms of 1D assessment metrics at the three Vigicrue observing stations, the free run (FR2) still significantly underestimates the water levels compared to the observations, especially around the flood peak at Marmande and La Réole as shown in \autoref{fig:VigicrueFR1FR2DA2}. The $\mathrm{RMSE}$ remains above 10 cm at the 3 stations, the $\mathrm{MaAE}$ found at the flood peak is close to 1.5 m at Marmande.
Similar conclusions are drawn for 2019 event, with a significant decrease of the $\mathrm{RMSE}$ at observing stations resulting from the bias correction and a residual $\mathrm{MaAE}$ of 65.4 cm at Marmande in FR2 (\autoref{tab:stats_2019}). 
 It is shown that the bias correction does not suffice to properly simulate the flood event variability and the flood peak. Thus DA assimilation should be applied in order to allow for a time varying correction of friction and inflow leading to improved simulation and forecast in the river bed and the flood plain.

 \begin{table*}[t] 
\centering
\caption{2D scores with respect to FloodML Sentinel-1 flood extent maps.}
\label{tab:scores}
\begin{minipage}{0.45\textwidth}
\begin{subtable}{\linewidth}
\centering
\caption{2D scores - 2021 flood event}
\label{stab:scores_2021}
\begin{tabular}{c|c|c|c|c}
\hline
& \multicolumn{4}{c}{Critical Success Index (\%)} \\ \hline
& 2021-01-28 & 2021-02-02 & 2021-02-03 & 2021-02-07 \\
& 19:00 & 19:00 & 19:00 & 07:00\\
\hline
{Free Run} & 25.09 & 44.39 & 55.86 & 22.10  \\ \hline
{DA1} & 23.58 & 41.91 & 61.97 & 20.00 \\ \hline
{DA2} & 24.95 & 43.98 & 63.84 & 20.87 \\ \hline
\end{tabular}
\end{subtable}

\vspace{0.05cm}

\begin{subtable}{\linewidth}
\centering
\begin{tabular}{c|c|c|c|c}
\hline
& \multicolumn{4}{c}{$F_1$-score (\%)} \\ \hline
& 2021-01-28 & 2021-02-02 & 2021-02-03 & 2021-02-07 \\
& 19:00 & 19:00 & 19:00 & 07:00\\
\hline
{Free Run} & 40.12 & 61.49 & 71.68 & 36.21  \\ \hline
{DA1} & 38.16 & 59.07 & 76.52 & 33.34 \\ \hline
{DA2} & 39.94 & 61.09 & 77.93 & 34.53 \\ \hline
\end{tabular}
\end{subtable}

\vspace{0.05cm}

\begin{subtable}{\linewidth}
\centering
\begin{tabular}{c|c|c|c|c}
\hline
& \multicolumn{4}{c}{Cohen's kappa index (\%)} \\ \hline
& 2021-01-28 & 2021-02-02 & 2021-02-03 & 2021-02-07 \\
& 19:00 & 19:00 & 19:00 & 07:00\\
\hline
{Free Run} & 39.11 & 57.81 & 67.82 & 33.73 \\ \hline
{DA1} & 37.09 & 54.54 & 72.82 & 30.68 \\ \hline
{DA2} & 38.92 & 57.25 & 74.58 & 31.94 \\ \hline
\end{tabular}
\end{subtable}
\end{minipage}
\hfill
\begin{minipage}{0.45\textwidth}
\begin{subtable}{\linewidth}
\centering
\caption{2D scores - 2019 flood event}
\label{stab:scores_2019}
\begin{tabular}{c|c|c|c|c}
\hline
& \multicolumn{4}{c}{Critical Success Index (\%)} \\ \hline
& 2019-12-15 & 2019-12-16 & 2019-12-17 & 2019-12-21 \\
& 07:00 & 19:00 & 19:00 & 07:00\\
\hline
{Free Run} & 30.44 & 42.95 & 30.15 & 16.48  \\ \hline
{DA1} & 20.45 & 44.75 & 28.66 & 12.87 \\ \hline
{DA2} & 27.77 & 47.74 & 31.55 & 14.23 \\ \hline
\end{tabular}
\end{subtable}\\

\vspace{0.05cm}

\begin{subtable}{\linewidth}
\centering
\begin{tabular}{c|c|c|c|c}
\hline
& \multicolumn{4}{c}{$F_1$-score (\%)} \\ \hline
& 2019-12-15 & 2019-12-16 & 2019-12-17 & 2019-12-21 \\
& 07:00 & 19:00 & 19:00 & 07:00\\
\hline
{Free Run} & 46.67 & 60.09 & 46.33 & 28.30  \\ \hline
{DA1} & 33.96 & 61.83 & 44.56 & 22.80 \\ \hline
{DA2} & 43.47 & 64.63 & 47.97 & 24.91 \\ \hline
\end{tabular}
\end{subtable}\\

\vspace{0.05cm}

\begin{subtable}{\linewidth}
\centering
\begin{tabular}{c|c|c|c|c}
\hline
& \multicolumn{4}{c}{Cohen's kappa index (\%)} \\ \hline
& 2019-12-15 & 2019-12-16 & 2019-12-17 & 2019-12-21 \\
& 07:00 & 19:00 & 19:00 & 07:00\\
\hline
{Free Run} & 44.03 & 55.99 & 43.52 & 26.70 \\ \hline
{DA1} & 30.34 & 57.32 & 41.32 & 21.01 \\ \hline
{DA2} & 40.59 & 60.74 & 45.07 & 23.19 \\ \hline
\end{tabular}
\end{subtable}
\end{minipage}
\end{table*}

\subsubsection{Flood extents assessment - Experiments FR1 and FR2}
The dynamic of the flood plain is assessed with respect to flood extents retrieved from Sentinel-1 data and FloodML algorithm. Nine flood extent maps were observed and generated during the 2021 event as indicated in \autoref{fig:all_obs_2021}. 
Four model output dates close to the flood peak are analyzed in the following (2021-01-28, 2021-02-02, 2021-02-03, and 2021-02-07). 
The $\mathrm{CSI}$, $F_1$-score and $\kappa$ scores (presented in \autoref{subsect:metrics}) are gathered in \autoref{stab:scores_2021} for FR1. FR2 yields similar results as no bias is estimated in the flood plain, thus FR1 and FR2 are presented as Free Run. As expected from previous analysis at in-situ observing stations, the free run simulation tends to underestimate the flood extent at all observed time steps.  For instance, at the  Sentinel overpass time closest to the flood peak (2021-02-03 18:48), the T2D flood extent ($H_\mathrm{FR1} > 0.05$ m) shown in green \autoref{fig:FML+FR1+OSO16_2021_FloodML}, significantly underestimates the FloodML flood extent (shown in the background binary map, where white pixels represent observed flood areas) especially in the flood plain upstream of Marmande. The under-predicted areas are color-coded in yellow in \autoref{sfig:FreeRun_1710000_2021_pred}, a large under-predicted area is located in the right bank flood plain upstream of Marmande and smaller areas under-predicted areas are located in the left bank flood plain near and downstream of Marmande. It should be noted that the correctly predicted flooded pixels are located in the flood plains, within distance from the river bank.  Additionally, the over-predicted pixels are mostly located in dense vegetation areas (shown in red in \autoref{fig:FML+FR1+OSO16_2021_FloodML}) where the FloodML maps is unreliable. Thus, these areas were removed from score computation and the resulting 2D metric scores are $\mathrm{CSI}$ = 55.86\%, $F_1$ = 71.68\% and $\kappa$ = 67.82\% on 2021-02-03 18:48. It should be noted that there could be small over-predicted areas found in locations where the land cover is not classified as deciduous forest but where the FloodML detection is still questionable (e.g. due to the presence of other vegetation class or urban areas). These potential areas are not excluded from the score computation in the present work.

For the 2019 event, four Sentinel-1 acquisitions close to the first peak are analyzed (2019-12-15, 2019-12-16, 2019-12-17, and 2019-12-21). Simulated flood extent and their associated contingency map resulted from FR1 compared to FloodML binary map are shown in \autoref{fig:FML+FR1+OSO16_2019_FloodML} and \autoref{sfig:FreeRun_414000_2019_pred}. The large under-predicted area (color-coded in yellow) in the right bank flood plain upstream of Marmande that was observed for 2021 remains, whereas large over-predicted areas appear in the flood plain upstream of La Réole. 
These over-predicted areas are not related to vegetation-impaired FloodML detection areas. Their causes are likely due to the possible inconsistency between the 2021-diagnosed bias and the dynamics of the 2019 flood event dynamic with large water level variations.
The resulting $\mathrm{CSI}$, $F_1$ and $\kappa$ scores for 2019 event are given in \autoref{stab:scores_2019}. 
For instance, at the closest Sentinel-1 overpass time to the first flood peak (2019-12-16 18:56), the resulting 2D metrics are $\mathrm{CSI}$ = 42.95\%, $F_1$ = 60.09\% and $\kappa$ = 55.99\%. These scores are lower than that computed for 2021 event, probably due to the presence of the large over-prediction area downstream of the catchment. 

\begin{figure*}[t]
    \centering
    \begin{minipage}{0.45\textwidth}
    \begin{subfigure}[b]{\linewidth}
        \centering
        \includegraphics[width=\linewidth]{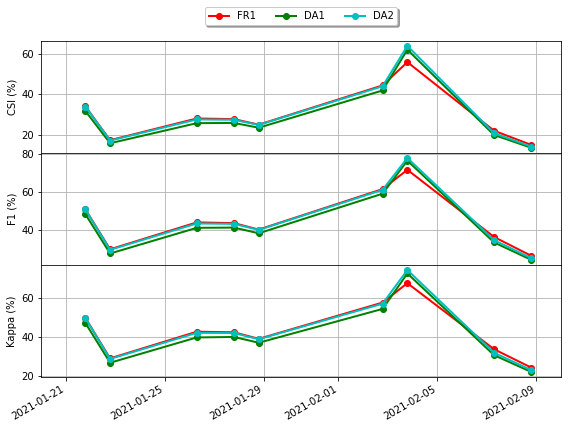}
        \caption{2021 flood event}
    \end{subfigure}
    \end{minipage}
    \hfill
    \begin{minipage}{0.45\textwidth}
    \begin{subfigure}[b]{\linewidth}
        \centering
        \includegraphics[width=\linewidth]{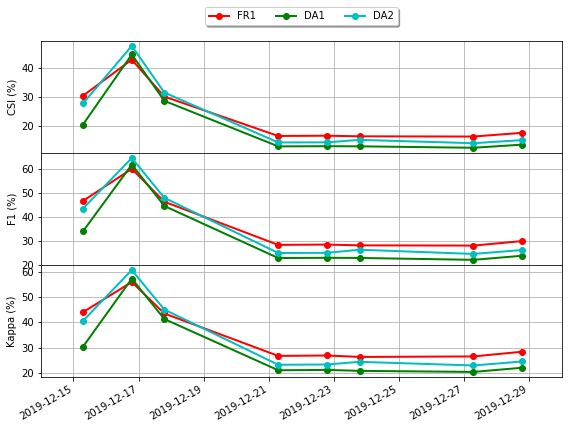}
        \caption{2019 flood event}
    \end{subfigure}
    \end{minipage}
    \caption{Comparison of flood extent maps in terms of 2D scores ($\mathrm{CSI}$, $F_1$-score and $\kappa$) for (a) 2021 flood event, (b) 2019 flood event.  FR1 is plotted in red, DA1 in green and DA2 in cyan.}
    \label{fig:timeseriesCSI}
\end{figure*}

Complete 2D metric scores, with respect to the Sentinel-1 derived independent data for the entire flood events at every Sentinel-1 overpass times are depicted in \autoref{fig:timeseriesCSI}, by red curves for the Free Run experiment (same results for FR1 and FR2). During the beginning of the events, the flow remains within the river bed and only occupies a small portion of the simulation domain, potentially not well observed by Sentinel-1 due to river banks and vegetation cover effects. Additionally, the numerical artificial flooding of the upstream first meander due to the prescription of the upstream BC (previously mentioned in \autoref{subsect:T2D}) penalizes the 2D scores in non-flooding conditions. Past the flood peak, these effects add up to the fact that T2D struggles to properly empty the flood plain when velocities are small as neither evaporation nor river-ground surface fluxes are taken into account in the T2D model, leaving still relatively significant volumes of water in low topography areas while they are drained or evaporated in the observation. 
Such a problem is more severe with the 2019 event with two flood peaks, which leads to very low 2D metrics (e.g. $\mathrm{CSI}$ under 20\%) after the first peak. The CSI score at the flood peak is close to 60\% for 2021 event and 40 \% for the 2019 event (first peak).

\subsubsection{Intermediate conclusion from Experiments FR1 and FR2}

The analysis of Free Run simulations allows to quantify a model-observation bias that leads to an under-estimation of water level in the river bed as well as of the flood extent in the flood plain. Comparison with independent RS data confirm the model behavior over the whole simulation domain. These results stand for both 2021 and 2019 flood events. A rough estimate of the bias was achieved over the beginning of 2021 event with FR1 and is further taken into account in the simulation FR2. This bias is coherent with the dynamics of the 2019 event (except at La Réole). Yet a time varying correction is needed to further improve the flood peak simulation as FR2 still underestimates the flood peak. Indeed, the friction value and inflow should be corrected with the assimilation of in-situ data, in order to improve the simulation of the flood in the river bed and in the flood plain.

\subsection{Merits of RS data for data assimilation assessment - Experiments DA1, DA2}
\label{ssec:merits_DA1_DA2}
Implementing EnKF, in-situ water level observations are assimilated in two experiments over 2021 flood event; DA1 where the diagnosed model-observation bias is not considered and DA2 where the bias is taken into account in the observation operator described in Eq.~\eqref{eq:ctlequivobsbias}. In the following, illustrations are provided for the mean of the DA ensemble members. 

\subsubsection{Water level at Vigicrue observing stations - Experiments FR1 and DA1}

In the context of in-situ only observation, the straightforward strategy is DA1 (carried out over 2021 and 2019 flood events). The DA1 experiment consists in assimilating observed water levels at Tonneins, Marmande and la Réole. The model-observation bias $\mathbf{y}_{bias}$  is not accounted for in the observation operator. This discrepancy is only corrected through the DA analysis on the friction and inflow within the control vector.
As such, the observation operator for DA1 is described in \autoref{eq:ctlequivobs}. DA1 simulated water level is plotted in green in the top panels of  \autoref{fig:VigicrueFR1FR2DA2_Tonneins_2021} at Tonneins, \autoref{fig:VigicrueFR1FR2DA2_Marmande_2021} at Marmande and \autoref{fig:VigicrueFR1FR2DA2_LaReole_2021} at La Réole, and the errors between DA1 water level and the observation are plotted in green in the respective bottom panels.  
For both 2019 and 2021 events, the EnKF succeeds in retrieving friction and inflow parameters such that the water level at the observing stations is really close to the observations.  The improvement with respect to FR1 is more significant for the 2021 event than for the 2019 event as the dynamics of the two-peak 2019 event (as well as the rapid rise of water level at the start) is more difficult to simulate with T2D. 

\subsubsection{$\mathrm{RMSE}$, maximum absolute error and $\mathrm{NSE}$ at Vigicrue observing stations - Experiments FR1 and DA1}

For the 2021 event, the $\mathrm{RMSE}$ computed over the flood duration (2021-01-15 to 2021-02-15) between DA1 water level and the observation are 9.0 cm, 5.9 cm and 14.8 cm at respectively Tonneins, Marmande and La Réole (cf. \autoref{tab:stats_2021}). These $\mathrm{RMSE}$ are reduced of 88.1\%, 90.6\% and 63.8\% with respect to those of FR1 at respectively Tonneins, Marmande and La Réole. The maximum absolute error and the $\mathrm{NSE}$ are also significantly reduced with respect to those of FR1, thus demonstrating the merits of data assimilation with respect to the reference run FR1. The maximum absolute error around the 2021 flood peak at Tonneins, and Marmande decreases from 1.06 m and 1.87 m for FR1 to less than 50 cm. For the 2019 event, the 1D metrics computed over the flood duration (2019-12-13 to 2021-12-29) are provided in \autoref{tab:stats_2019}, the $\mathrm{RMSE}$ is reduced of 88.4\%, 90.0\% and 2.5\% with respect to those of FR1 at respectively Tonneins, Marmande and La Réole. It should be noted that these results are similar to those for 2021 flood event, except at La Réole where the dynamics of the downstream flood plain remains incoherent with that of the upstream flood plain as illustrated by the over-prediction upstream of La Réole in \autoref{sfig:FreeRun_414000_2019_pred}. \\

\subsubsection{Control vector analysis - Experiments DA1 and DA2}

\begin{figure*}
    \centering
    \begin{minipage}{0.48\textwidth}
     \begin{subfigure}[b]{\textwidth}
         \centering
         \includegraphics[width=\textwidth]{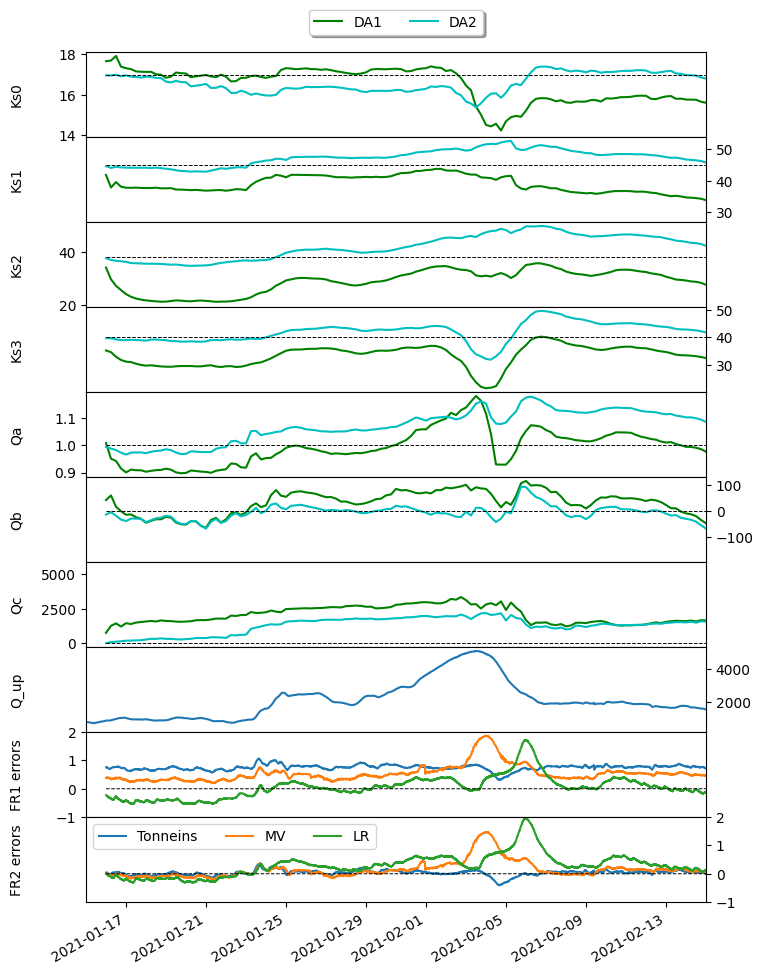}
         \caption{2021 flood event}
         \label{fig:param_compare_plot_2021}
     \end{subfigure}
     \end{minipage}
     \hfill
     \begin{minipage}{0.48\textwidth}
     \begin{subfigure}[b]{\textwidth}
         \centering
         \includegraphics[width=\textwidth]{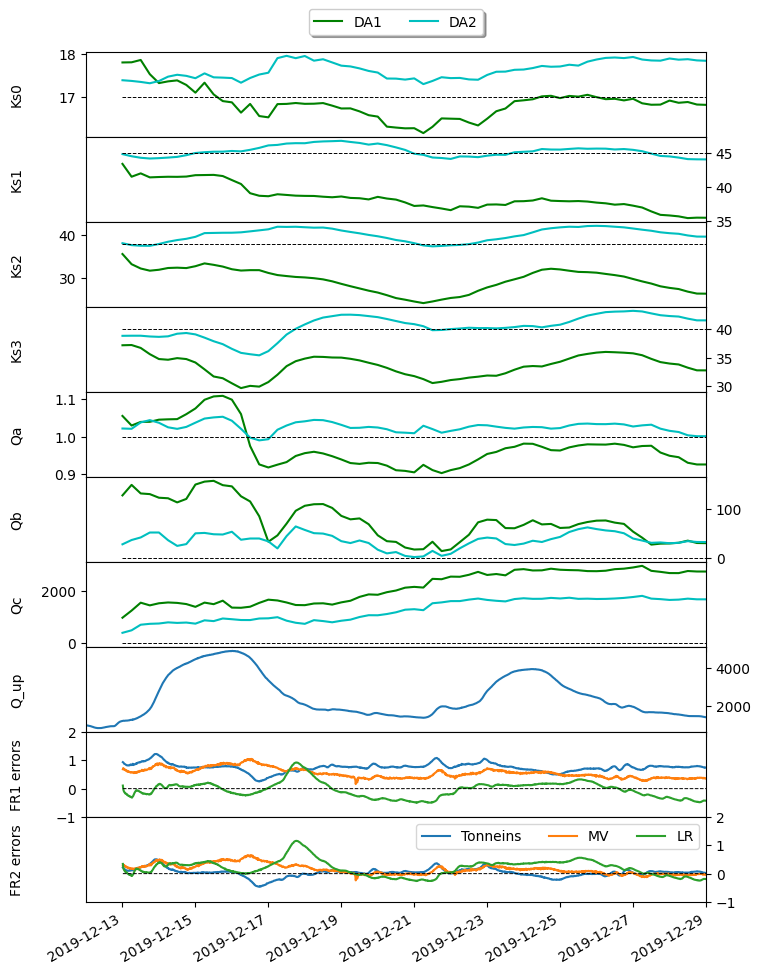}
         \caption{2019 flood event}
         \label{fig:param_compare_plot_2019}
     \end{subfigure}
     \end{minipage}
     \caption{Evolution of the control parameters for the 7 top panels ($ K_{s_0}$, $K_{s_1}$, $K_{s_2}$, $K_{s_3}, a, b, c $ over the flood events for DA1 (green curves) and DA2 (cyan curves) experiments, with respect to their background values (black dashed lines). The 3 bottom panels display, the upstream inflow $Q_{up}$ at Tonneins, the water level difference between FR1 (respectively FR2) and the observation at Tonneins (blue curve), Marmande (orange curve) and La Réole (green curve). (a) 2021 flood event and (b) 2019 flood event.}
     \label{fig:param_compare_plot}
\end{figure*}

The analyzed values from DA1 and DA2 for friction and inflow parameters are shown in green in the first top 7 panels (for 7 scalars) in \autoref{fig:param_compare_plot_2021} and \autoref{fig:param_compare_plot_2019} for the 2021 and 2019 flood events, respectively. 
Control parameters in DA1 are depicted by green curves and those in DA2 by cyan curves (same color-codes as previous \autoref{fig:VigicrueFR1FR2DA2} and \autoref{fig:timeseriesCSI}).
The initial background values for the flood plain friction coefficient $K_{s_0}$, river bed friction coefficients $K_{s_1}, K_{s_2}, K_{s_3}$ and inflow parametric corrective coefficients $a, b, c$ are indicated by the horizontal black dashed lines. The 3 bottom panels on each sub-figure display the upstream inflow at Tonneins (as an indication of the flood dynamics), the differences between FR1 (or FR2) water level and the observation at Tonneins (blue curve), Marmande (orange curve) and La Réole (green curve). 

For both 2021 and 2019 events, the EnKF algorithms computes analyzed friction and inflow values in order to correct the model-observation misfit, taking into account uncertainty in both sources (model and observation). Since the model-observation bias is not taken into account in the observation operator in DA1, as shown by the significant FR1 errors (ninth panel), this requires a large correction in the control vector for most of the control vector elements since the very beginning of the event. This correction intensifies to account for larger errors as the flood begins (2021-01-23 and 2019-12-14) and around the flood peak (2021-02-03, 2019-12-16 and 2019-12-23).
In other words, more efforts are required on all 7 parameters of the control vector for DA1 compared to DA2 to reduce the distance to the observations. This is indicated by the larger distance from the green and cyan curves to the background values on each parameter plots.
On the other hand, when the water level errors are small for DA2, e.g. between 2021-01-15 and 2021-01-21 (as shown in the tenth panel between FR2 water level and observation), thanks to the applied model-observation bias, the control vector parameters remains close to the background values.
Globally, the EnKF tends to increase the volume of water within the simulation domain by decreasing the Strickler friction coefficients and increasing the inflow hyper parameters (with respect to the background values), especially near the flood peak time. This allows to increase the water level at the observing stations and reach a better agreement with the observed rating curve at Tonneins. It should be noted that the resolution of the inverse problem with the EnKF may suffer from an equifinality issue and that the analyzed values of the control may compensate in order to lead to expected water level values; this will be further discussed in \autoref{sec:forecast} dedicated to the DA results in forecast mode.

\begin{figure}[h]
    \centering
    \includegraphics[width=\linewidth]{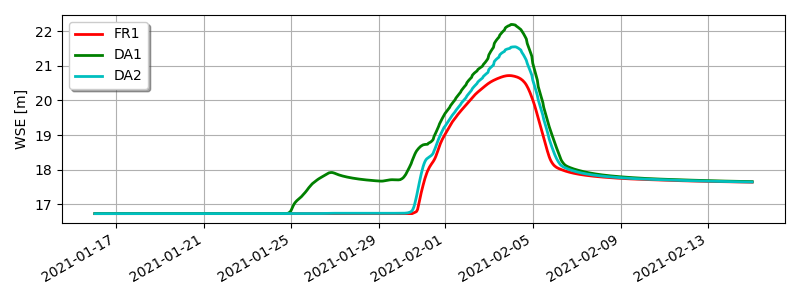}
         \caption{Water surface elevation at validation point Flood Plain near Marmande (FPM)}
         \label{fig:ZMarmandeFloodPlain}
\end{figure}

\begin{figure*}[!t]
     \centering
     \begin{minipage}{0.45\textwidth}
     \begin{subfigure}[b]{\linewidth}
        \centering
        \includegraphics[trim=1cm 0 0 0, clip, width=\linewidth]{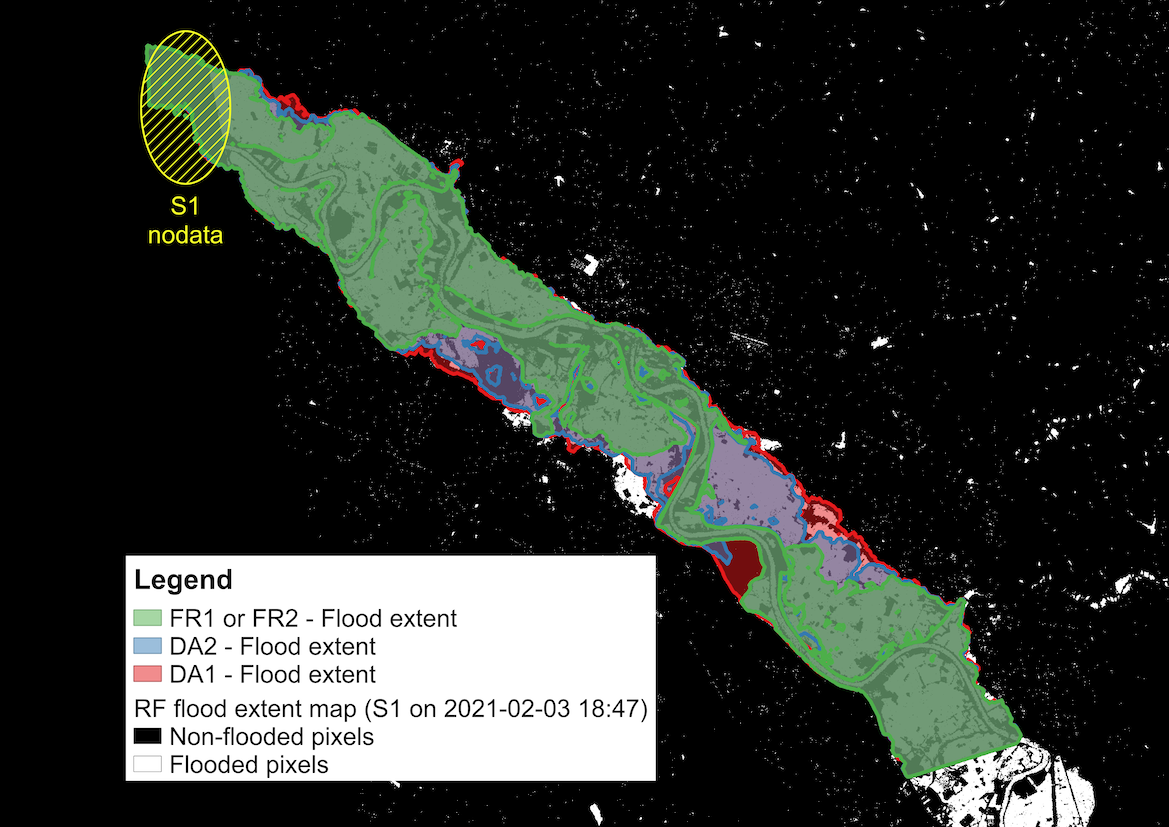}
         \caption{}
         \label{fig:FloodMLDA_2021}
     \end{subfigure}
     \begin{subfigure}[b]{\linewidth}
         \centering
         \includegraphics[width=\linewidth]{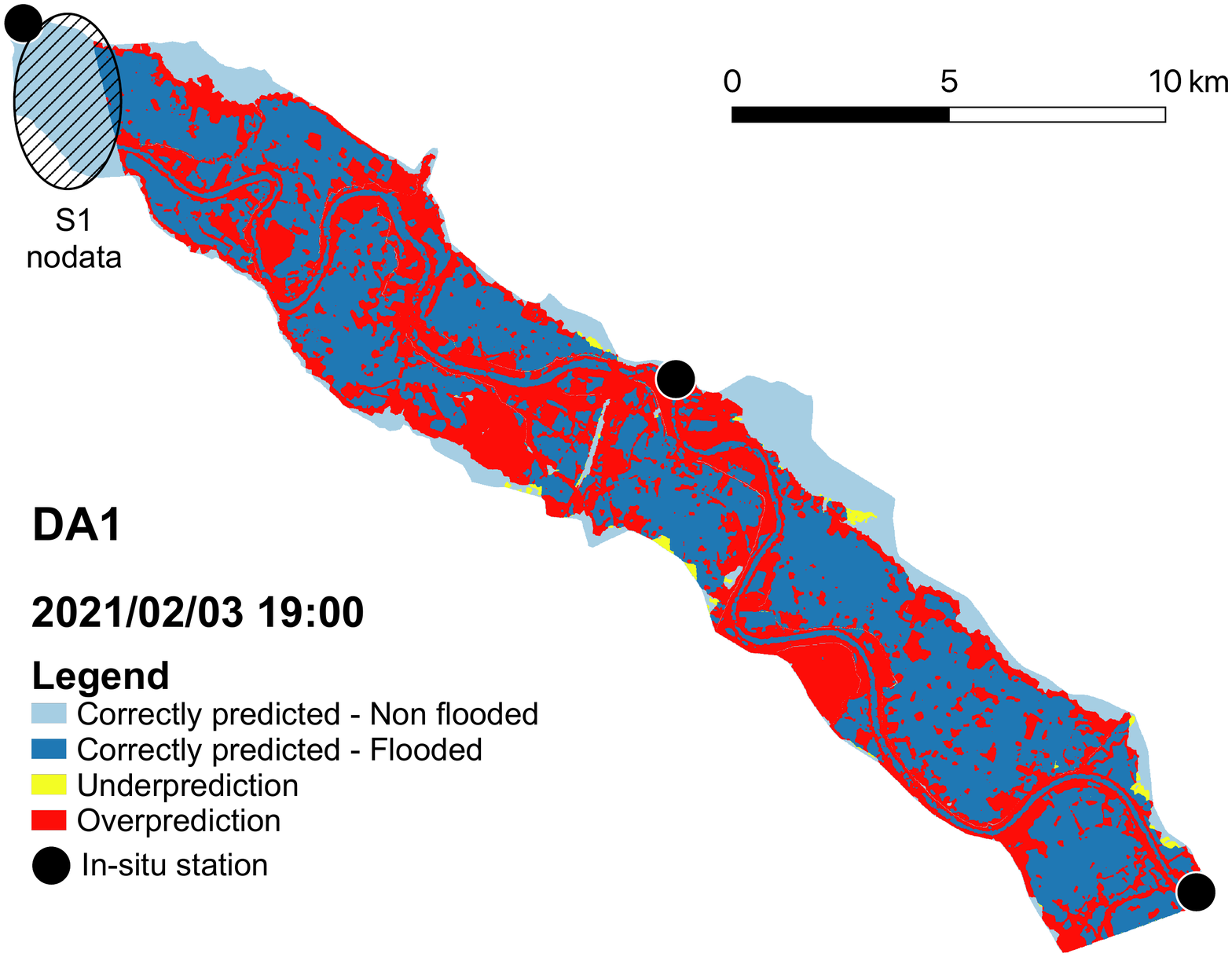}
         \caption{}
         \label{sfig:EnKF_1710000_DA1_2021}
     \end{subfigure}
     \begin{subfigure}[b]{\linewidth}
         \centering
         \includegraphics[width=\linewidth]{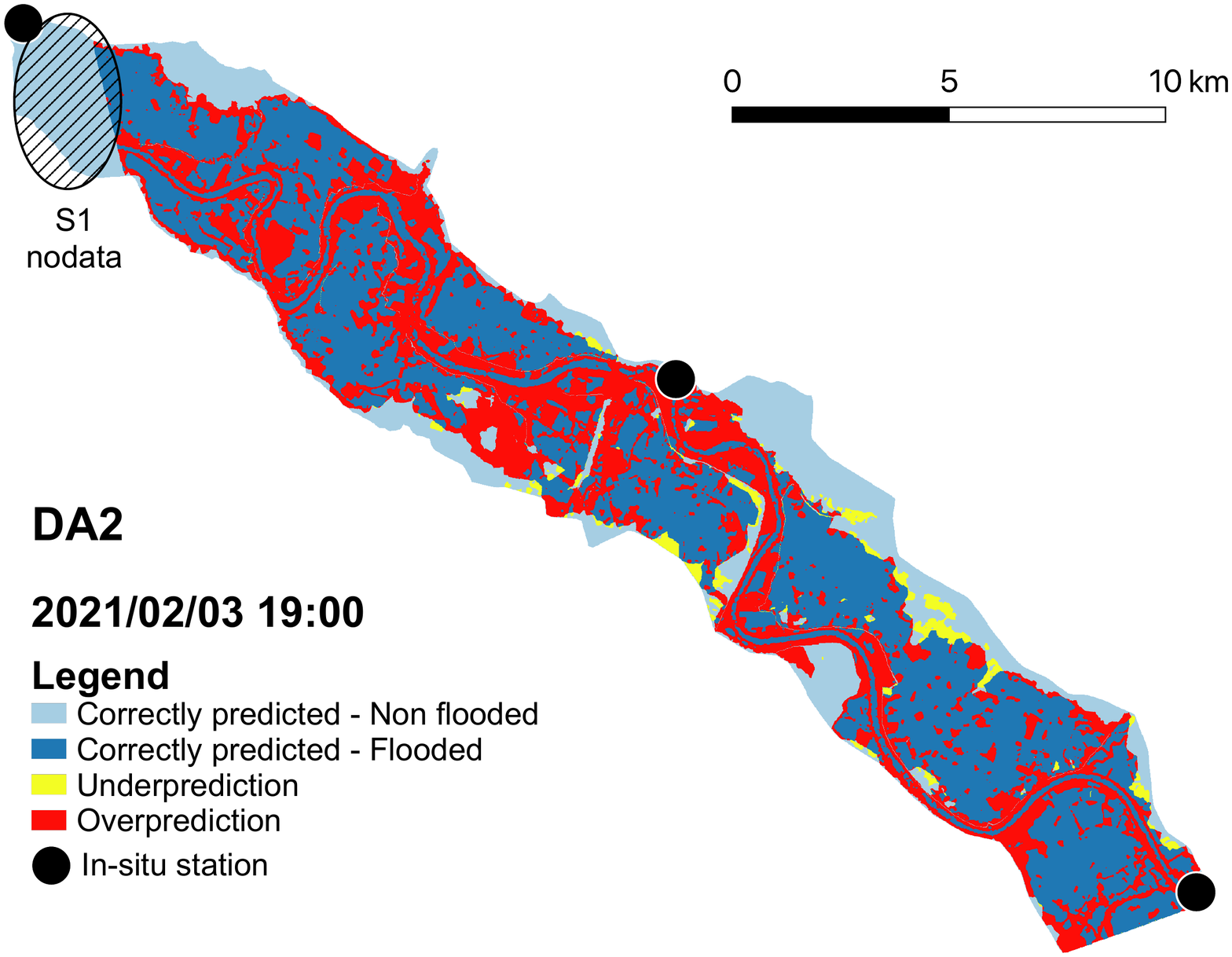}
         \caption{}
         \label{sfig:EnKF_1710000_DA2_2021}
     \end{subfigure}
     \end{minipage}
     \hspace{0.1cm}
     \begin{minipage}{0.45\textwidth}
      \begin{subfigure}[b]{\linewidth}
        \centering
        \includegraphics[trim=1cm 0 0 0, clip, width=\linewidth]{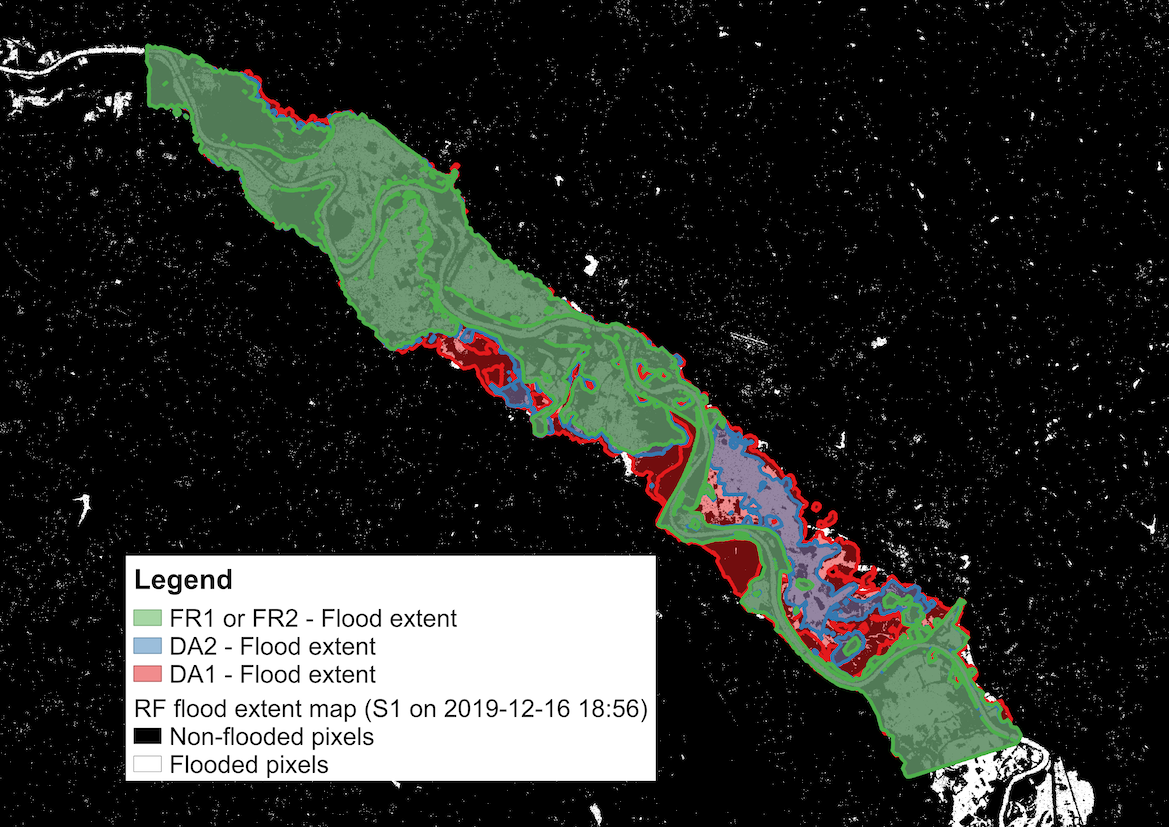}
         \caption{}
         \label{fig:FloodMLDA_2019}
     \end{subfigure}
     \begin{subfigure}[b]{\linewidth}
         \centering
         \includegraphics[width=\linewidth]{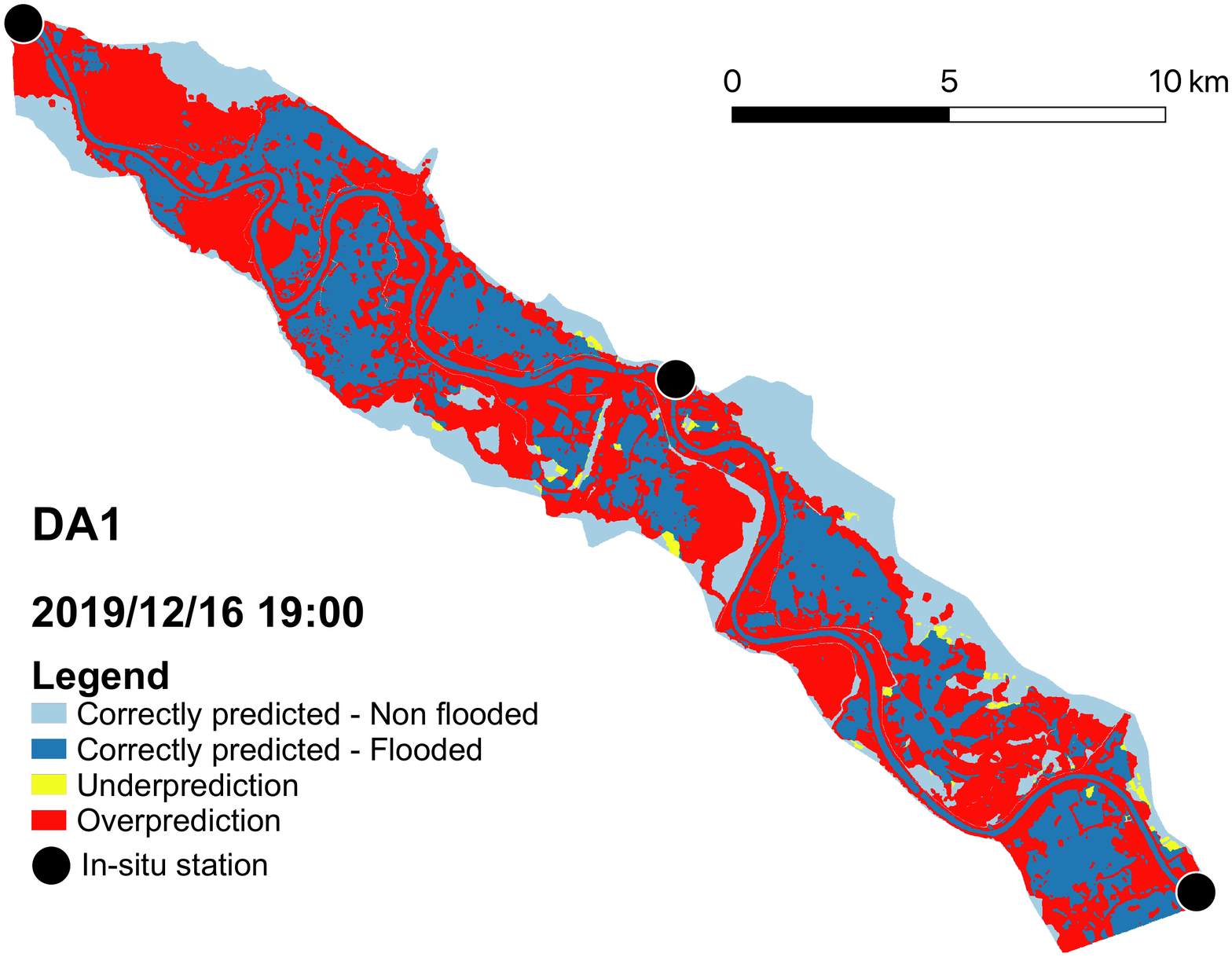}
         \caption{}
         \label{sfig:EnKF_414000_DA1_2019}
     \end{subfigure}
     \begin{subfigure}[b]{\linewidth}
         \centering
         \includegraphics[width=\linewidth]{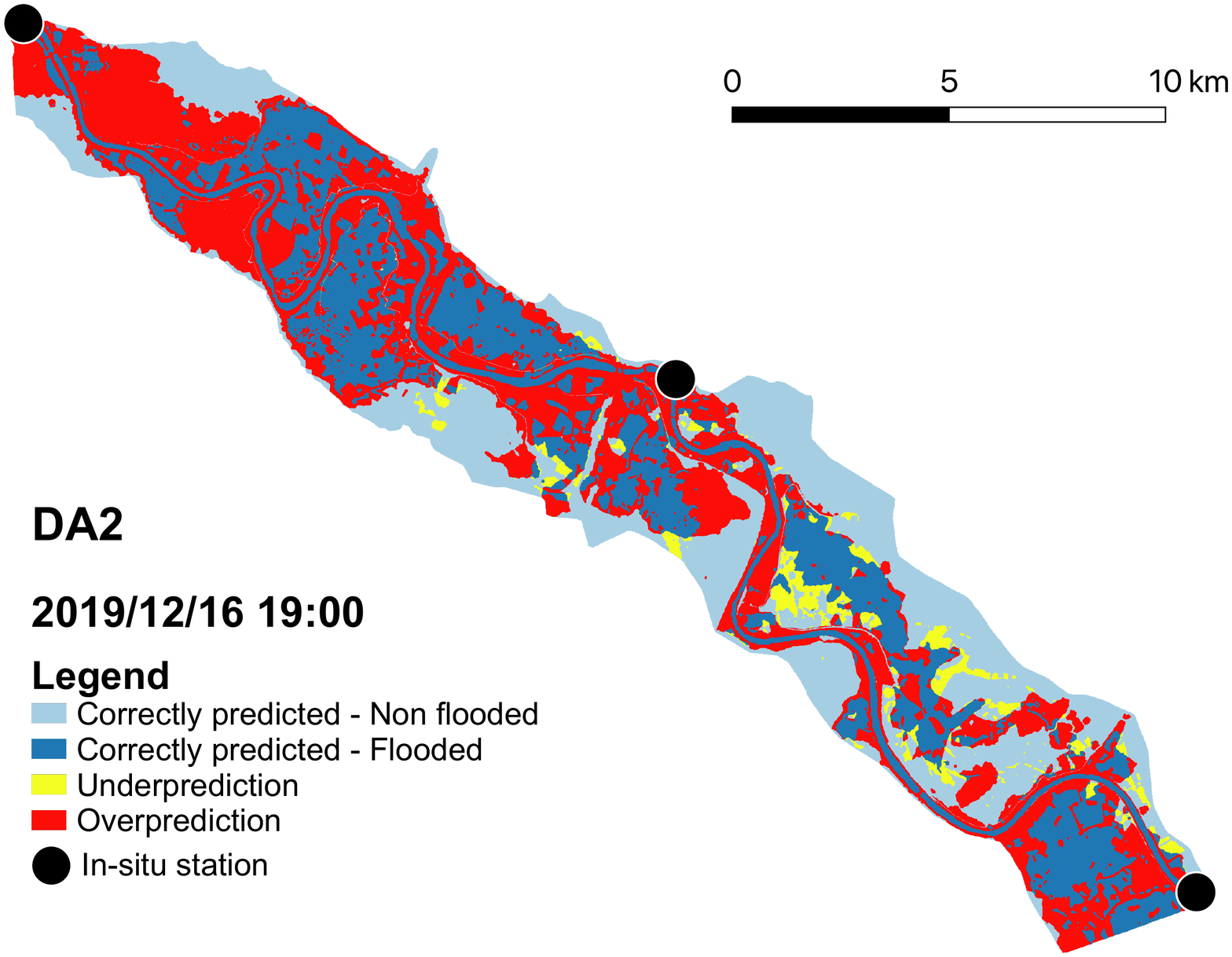}
         \caption{}
         \label{sfig:EnKF_414000_DA2_2019}
     \end{subfigure}
      \end{minipage}
     \caption{Comparison between FloodML and T2D flood extent maps. Left column: 2021-02-03, and right column: 2019-12-16. (a) Flood extent of the free run (green), DA1 (red outline), DA2 (blue outline) simulations overlapping on the Sentinel-1 flood binary map for 2021-02-03 (respectively, (d) for 2019-12-16). Contingency map representing the flood prediction by (b) DA1 and (c) DA2 experiments for 2021-02-03 (respectively, (e) and (f) for 2019-12-16). The correctly predicted flooded areas are represented in dark blue, correctly predicted non-flooded areas in light blue, under predicted areas in yellow, over predicted areas in red.
     }
     \label{fig:flood_extent_comp}
\end{figure*}

\subsubsection{Flood extents assessment - Experiments FR1 and DA1}

The comparison of DA1 results with independent RS flood extent highlights that while DA improves the water level results at the Vigicrue observing stations, it is at the expense of the flood plain dynamics. This is illustrated in \autoref{fig:ZMarmandeFloodPlain} which compares the simulated WSE at a selected location FPM (Flood plain near Marmande), i.e. the white solid circle shown in \autoref{fig:study_area}. It is worth-noting that there is no in-situ observation at this location.
While this location is not flooded in the free run simulation (red curve) before 2021-01-30 as shown by the WSE remaining constant, it is conversely flooded in DA1 experiment (green curve) from 2021-01-25. 

The use of RS flood extent provides complete and independent data to validate the dynamic of the flood plains for DA1 for 2021 and 2019 flood events as shown in \autoref{fig:flood_extent_comp} (left column for 2021-02-03, right column for 2019-12-16).
The T2D flood extents simulated with FR1 (respectively with DA1) are plotted in green (respectively in red), and the flood extent derived from Sentinel-1 image is represented by the background binary map in white pixels means flooded areas (similarly to \autoref{fig:FloodML}). 
It is revealed  in \autoref{fig:FloodMLDA_2021} and \autoref{sfig:EnKF_1710000_DA1_2021} for 2021 and in \autoref{fig:FloodMLDA_2019} and \autoref{sfig:EnKF_414000_DA1_2019} for 2019 that DA1 leads to the over-flooding of the domain over the entire catchment on both left and right river banks and flood plains.

In terms of quantitative assessments, the resulting $\mathrm{CSI}$, $F_1$-score and $\kappa$ indices are given in \autoref{stab:scores_2021} and \autoref{stab:scores_2019}, they are improved at the flood peak with respect to FR experiments (FR1 and FR2 have the same CSI scores by definition).  For instance, at the closest Sentinel-1 overpass time to the flood peak (2021-02-03 18:48), $\mathrm{CSI}$, = 61.97\% (compared to $\mathrm{CSI}_\mathrm{FR1}$ = 55.86\%), $F_1$ = 76.52\% (compared to $F_{1,\mathrm{FR1}}$ = 71.68\%), and $\kappa$ = 72.82\% (compared to $\kappa_\mathrm{FR1}$ = 67.82\%). 
Full 2D metric scores at all Sentinel-1 observation times plotted in \autoref{fig:timeseriesCSI} show an improvement with respect to FR1 for 2021. The results are slightly degraded for 2019. There are two main potential reasons for this: first the dynamics after the first flood peak is hard to simulate with T2D, then the limited size of the DA control vector does not allow for a finely tuned correction that is needed for this two-peaks event, especially as it should overcome a large model-observation misfit $\mathbf{y}_{bias}$.

\subsubsection{Merits of data assimilation when bias is removed - Experiments FR2 and DA2}

When the model-observation bias is taken into account in the observation operator (Eq. \eqref{eq:ctlequivobsbias}) in experiment DA2, the results of DA are greatly improved, first with respect to FR2, but also with respect to DA1. At the beginning of the 2021 event, the error between the model FR2 and the observation is small at all three observing stations (\autoref{fig:VigicrueFR1FR2DA2_Tonneins_2021}, \autoref{fig:VigicrueFR1FR2DA2_Marmande_2021}, \autoref{fig:VigicrueFR1FR2DA2_LaReole_2021}), thus the DA2 correction to the control vector (cyan curve) is significantly smaller then that of DA1 (green curve) as shown in \autoref{fig:param_compare_plot_2021} and \autoref{fig:param_compare_plot_2019} for 2021 and 2019 event respectively. It should be noted that this is a positive feature of DA2 as the EnKF algorithm is dedicated to weakly nonlinear problem only, thus more likely to provide an optimal solution when the size of the increment is limited \cite{asch2016data}.
For both events, a stronger correction occurs around the flood peak, as the nature of the errors evolves when the flood occurs. Same results holds for 2019 event, yet, the error between model FR2 and the observation is larger than that for FR1 (as the bias is diagnosed based on 2021 event), thus the DA2 correction made to the control vector is larger than that of 2021.

In terms of 1D assessment metrics, DA2 water level is close to the observation over the entire event with $\mathrm{RMSE}$, maximum absolute error and $\mathrm{NSE}$ (\autoref{tab:stats_2019} and \autoref{tab:stats_2021} close to that of DA1, and significantly smaller than that of FR1 and FR2. While DA2 simulated water level is not as good as that of DA1 at the flood peak, DA2 yields a significantly better dynamics in the flood plain than DA1 as shown in \autoref{sfig:EnKF_1710000_DA2_2021} and \autoref{sfig:EnKF_414000_DA2_2019}. The over-predicted areas from DA1 are no longer present and the flood extents at 2021-02-03  and 2019-12-16 are in good agreement with the Sentinel-1 derived flood extents. Resulting 2D metrics are improved compared to DA1 and the Free Run, as such $\mathrm{CSI}$ = 63.84\%, $F_1$ = 77.93\%, and $\kappa$ = 74.58\% (cf. \autoref{tab:scores}). This improvement stands for the flood peak simulation time for 2019 event. 
\autoref{fig:timeseriesCSI} shows that DA2 (cyan curve) has better 2D scores than DA1 (green curve) for all Sentinel-1 acquisition times in 2021 and 2019.
It should also be noted in \autoref{fig:ZMarmandeFloodPlain} that at diagnosed location FPM, the WSE from DA2 (cyan curve) no longer artificially increases between 2021-01-25 and 2021-01-30 (thus not flooding this location) as seen in DA1 (green curve).

In conclusion, the use of SAR-derived flood extent allows to conclude that DA2 provides the best strategy to correct the model friction and the inflow so as to improve the water level at the Vigicrue in-situ observing stations as well as in the flood plains (both water level and flood extent). DA2 is thus further studied in forecast mode in the following \autoref{sec:forecast} for lead time up to +24h.

\subsection{Forecast results DA2}\label{sec:forecast}

\begin{figure}[!ht]
 \centering
    \begin{subfigure}[b]{\linewidth}
    \centering
    \includegraphics[width=\textwidth]{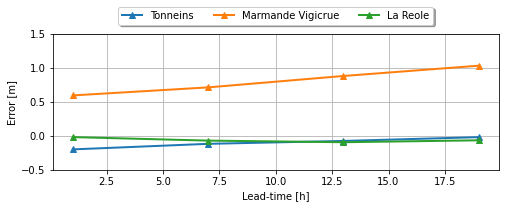}
    \caption{Target date: 2021-02-03 19:00}
    \label{fig:DA2_fct_targetdate_2021}
    \end{subfigure}
    \begin{subfigure}[b]{\linewidth}
    \centering
    \includegraphics[width=\textwidth]{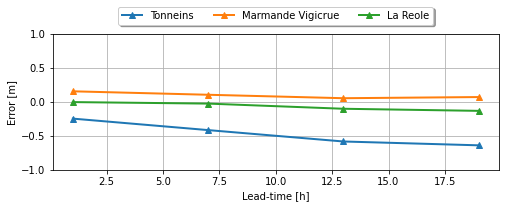}
    \caption{Target date: 2019-12-16 19:00}
    \label{fig:DA2_fct_targetdate_2019}
    \end{subfigure}
    \caption{Forecast error at the target dates 2021-02-03 19:00 and 2019-12-16 19:00 with various lead-times.}
\end{figure}

 \begin{figure*}[!t]
 \centering
    \begin{minipage}{0.45\textwidth}
    \begin{subfigure}[b]{\linewidth}
    \centering
    \includegraphics[width=\textwidth]{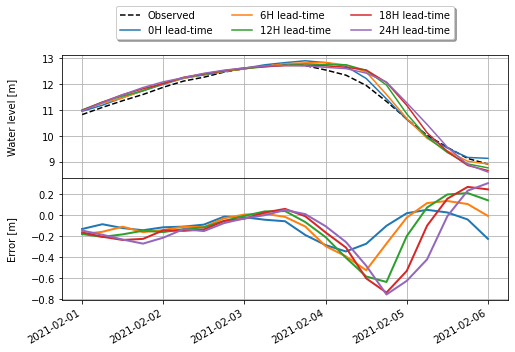}
    \caption{Tonneins}
    \end{subfigure}
    \begin{subfigure}[b]{\linewidth}
    \centering
    \includegraphics[width=\textwidth]{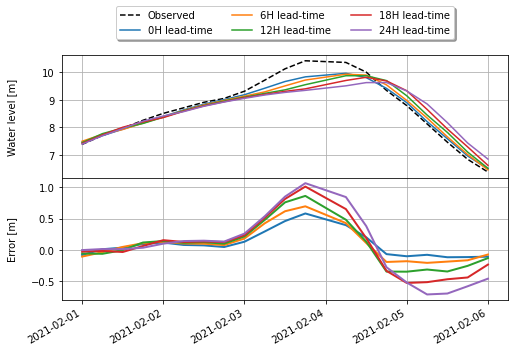}
    \caption{Marmande}
    \end{subfigure}
    \begin{subfigure}[b]{\linewidth}
    \centering
    \includegraphics[width=\textwidth]{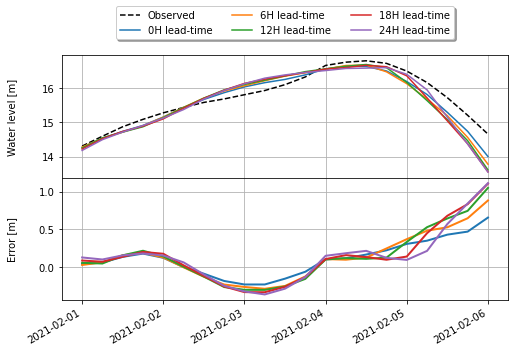}
    \caption{La Réole}
    \end{subfigure}
    \end{minipage}
    \hfill
    \begin{minipage}{0.45\textwidth}
    \begin{subfigure}[b]{\linewidth}
    \centering
    \includegraphics[width=\textwidth]{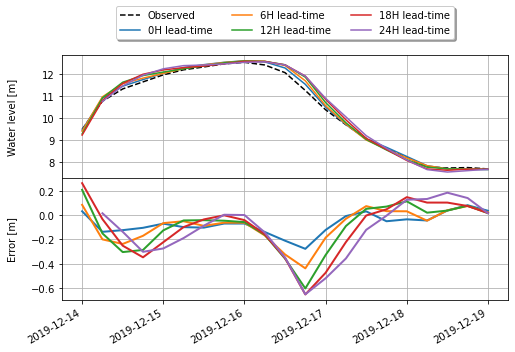}
    \caption{Tonneins}
    \end{subfigure}
    \begin{subfigure}[b]{\linewidth}
    \centering
    \includegraphics[width=\textwidth]{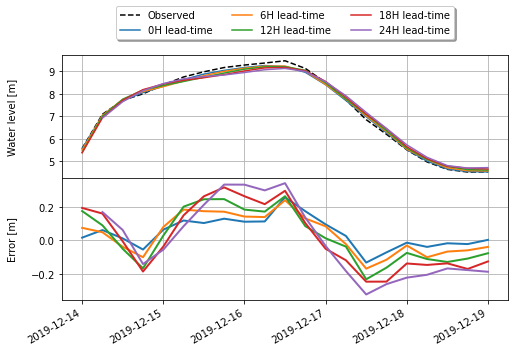}
    \caption{Marmande}
    \end{subfigure}
    \begin{subfigure}[b]{\linewidth}
    \centering
    \includegraphics[width=\textwidth]{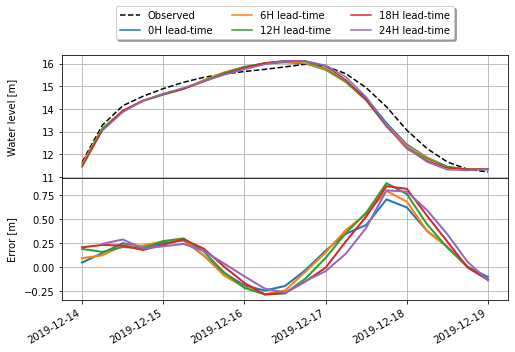}
    \caption{La Réole}
    \end{subfigure}
    \end{minipage}
     \caption{Forecast DA2 water levels at increasing lead times (+6h, +12h, +18h, +24h) and their errors with respect to the observed water levels at Vigicrue observing stations (a) Tonneins, (b) Marmande, and (c) La Réole for 2021 flood event (respectively, (d)-(f) for 2019 flood event).}
    \label{fig:DA2_fct_leadtime}
\end{figure*}

The forecast issued from DA2 are assessed here, focusing on the ensemble forecast mean. 
DA2 results at the flood peak in forecast mode are shown in \autoref{fig:DA2_fct_targetdate_2021} and \autoref{fig:DA2_fct_targetdate_2019} for the target date 2021-02-03 19:00 and 2019-12-16 19:00 respectively for lead time increasing from +1h to +7h, +13h and +19h ($x$-axis). 
The difference between DA2 water level and the observation is plotted along the $y$-axis for Tonneins (blue curve), Marmande (orange curve) and La Réole (green curve). 
This diagnostics implies the post-processing of four 12-hour DA cycles that provide an updated forecast every 6h for the target date. 
It appears that, as expected, the quality of the forecast mostly decreases  as the lead time increases. The degradation of the forecast is most evident at Marmande, with a positive error (under-estimation) for 2021 peak and at Tonneins with a negative error (over-estimation) for 2019 peak. The quality of the other forecast remain stable, this leads to the satisfying conclusion that the nature of the error between model and observation is stationary over the assimilation and the forecast period for the considered target dates; thus the forecast strategy is reliable.

\begin{figure}[!t]
 \centering
    \begin{subfigure}[b]{\linewidth}
    \centering
    \includegraphics[width=\textwidth]{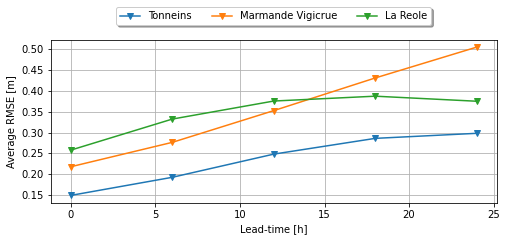}
    \caption{2021 flood event}
    \end{subfigure}
    \begin{subfigure}[b]{\linewidth}
    \centering
    \includegraphics[width=\textwidth]{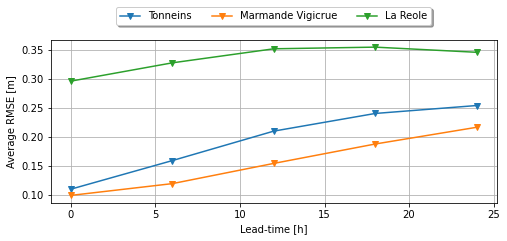}
    \caption{2019 flood event}
    \end{subfigure}
    \caption{$\mathrm{RMSE}$ between DA2 forecast water level and observation, computed  over the flood peak period, (a) 2021-02-01 to 2021-02-06 and (b) 2019-12-14 to 2019-12-19, at Tonneins (blue curve), Marmande (orange curve) and Lé Réole (green curve), along increasing lead time.}
    \label{fig:RMSE_fct}
\end{figure}

The forecast capability of the DA strategy is assessed over a couple of days around flood peak in 2021 (left column) and 2019 (right column) event for given lead time in \autoref{fig:DA2_fct_leadtime} at Tonneins, Marmande and La Réole. Colored solid curves in the top panels indicated the forecast water level with increasing lead time (+0h in blue, +6h in orange, +12h in green, +18h in red, +24h in purple) and the observed water level is plotted with a black dashed curve. The same color code is kept to represent the error between the forecast and the observation for the bottom panels. It is shown that the simulated flood peak is delayed and flatten and that the error (with respect to the observation) amplifies as the lead time increases, leading to up to $1$ m error at Marmande in 2021 (coherent with conclusion for \autoref{fig:DA2_fct_targetdate_2021}).  It should be noted that the sensitivity to the forecast lead time is less significant at La Réole. These results are synthetically presented by the $\mathrm{RMSE}$ between DA2 forecast and the observation, computed over a couple of days around the flood peak, depicted in \autoref{fig:RMSE_fct} for each observing station (Tonneins with a blue curve, Marmande with an orange curve and la Réole with a green curve) for increasing lead times. It clearly appears that around the flood peak, the quality of the forecast decreases as the forecast lead time increases, yet remaining below $50$ cm. This advocates for a fine resolution of the DA cycling, and frequent forecast updates.

 \section{Discussions}\label{sect:Discussions}
 
 DA allows to improve water level simulation and forecast in the river bed where observing stations are located as well as in the flood plain when the model-observation bias is taken into account in the observation operator. This validation was made possible by the use of RS flood extent derived from Sentinel-1 images. 
 Indeed, the involvement of RS flood extent advocates for the use of spatially distributed data, overcoming the limits of existing calibration and DA process. 
 Yet, the DA results remain imperfect and some under- and over-prediction areas are observed. Alternative settings for the DA algorithms are presented in \cite{NguyenTUC2021} with the focus on observation error estimation $\sigma_{obs}$. Two hypotheses to further improve the results are discussed here: the definition of the control vector for DA and the a priori estimate of the model-observation bias.
 
On the one hand, the definition of the control space should be coherent with the nature of the model-observation error over time and space. In the present study, it is most likely that the present control vector should be extented. Indeed, in FR1 in \autoref{sfig:FreeRun_414000_2019_pred}, for 2019 floodpeak, there is an over-prediction area before La Réole while the flood is properly represented after Marmande. As Marmande and La Réole are located in the same friction zone ($K_{s_3}$), DA fails to account for these opposite errors by tuning the control vector. Same conclusion holds for the upstream inflow correction, all the more as there is an under-prediction area upstream of Marmande for both events. The size of the control vector should be increased to a larger number of friction areas in the river bed and in the flood plain in order to allow for a refined correction. The extension of the control vector to an lateral inflow to the model could also be investigated in order to input more water in the system, for instance near the flood peak, when tributaries may bring in significant amount of water. 

On the other hand, the estimation of the model-observation bias is a key step in the DA strategy as it allows for smaller correction and more optimal analysis. Yet, this estimation was made roughly here on the beginning of 2021 event during a stationary non-dry period. Other estimation strategies could be investigated, for instance, considering permanent flow simulations would allow to diagnose a bias with a dependency on the discharge in the river. The bias could also be estimated over a large number of observed and simulated flood events. Finally, as the model error can vary over time, it could be included in the control vector and and corrected by the DA algorithm.

\section{Conclusions, Limitations and Perspectives}\label{sect:ConcPers}

In this paper, flood extents extracted from Sentinel-1 images using Machine Learning algorithms (Random Forest) were used to assess the performance of an ensemble-based data assimilation algorithm that assimilates in-situ water level and corrects friction coefficients and inflow discharge. The study is carried out over the Garonne Marmandaise catchment with Telemac2D SWE numerical solver, focusing on two significant and recent flood events in 2019 and 2021. The data assimilation is implemented as an ensemble Kalman Filter with a 12h-sliding assimilation window, and a 6h-overlap. An ensemble of forecast is issued from each analyzed state maintaining the updated friction and inflow correction over 24h. Four experiments are carried out, two in free run mode (FR1 and FR2) and two in DA mode (DA1 and DA2). A model-observation bias was diagnosed over the beginning of 2021 event in FR1; it is removed in the observation operator in FR2 and DA2.  
T2D simulation results are compared to in-situ and RS-derived observations as well as to deterministic free run simulation, with 1D and 2D metrics. 

It was shown that when the model-observation bias is not taken into account in the observation operator, the assimilation DA1 fails to recover the dynamics of the river bed and the flood plain. The water level at the in-situ observing station is greatly improved while the flood plain is over-flooded and the DA increments are large. When the bias is accounted for in the DA algorithm (i.e. DA2), the water level is properly corrected in both the river bed and the flood plain, without artificial flooding. The largest correction are found around the flood peak period. These diagnoses were only made possible by the use of a densified observing network with RS-derived flood extents; this constitutes an innovative piece of work. The merits of working with independent 2D flood extent data were demonstrated in re-analysis mode; the model was improved with DA thus leading to improved forecasts. It was shown that the quality of the forecast tends to decrease as the lead time increases, while remaining below 50 cm for +24h lead time. This demonstrates the forecast capability of the DA strategy and advocates for the use of updated forecasts from the cycled DA, eventually lowering the cycling overlap near the flood peak.

Some limitations were noted in the SAR-derived flood extent maps, mostly due to the dense vegetation coverage which causes mis-detection of flooded areas. These areas were identified with IOTA2 land cover maps and excluded from 2D assessment metric computation. 
Another drawback is due to the classification accuracy of Random Forest algorithm which is a reasonable solution for rapid flood mapping but could be outperformed (in terms of overall accuracy) by more recent Deep Learning classification methods.
Furthermore, some limitations were also noted in the simulation of the flood recession with T2D as evaporation and ground infiltration physical processes are not accounted for in the Garonne model. Therefore, the flood extent validation was mostly carried out during the flood rise and the flood peak, leaving aside post-peak flood extent observation. Further works stand in the constitution of an exclusion mask in the simulated and observed water extents comparison step, in order to deal with the presence of permanent waters, main river channel, urban and vegetation areas.
The major shortcoming for the DA algorithm stands in the limitation of the control vector size that does not allow for a space-varying correction finer than that of the friction coefficient uniform definition in 4 zones (and only 1 zone for the whole flood plain). An a priori finer zoning of the friction could follow the soil physical composition and vegetation classification. Finally, the control vector could also include lateral inflow to the river/flood plain network to account for tributaries that may become significant for high flow dynamics. It should be noted that the choice of the DA algorithm---eventually along with the ensemble size---may need to be revisited as the control vector is augmented and the assumption on the errors statistics reconsidered.
 
A major perspective for this study stands in the assimilation of Sentinel-1 derived flood-related information (e.g. flood extent, flood depth) in complement with the assimilation of in-situ observations. \cite{Schumann2008a,Schumann2008b,Blosch2001} asserted that the densification of the observing network lowers the equifinality issue and lead to improved forecasts. The merits of assimilating in-situ data in the flood plains were demonstrated in \cite{Wesemael2018, Ziliani2019}. \cite{hostache2009water} further stated that the local nature of field data can lead to an over-fitting of the hydraulic model to a specific location or event, thus limiting the overall predictive skills of the model. The assimilation of flood extent information will allow for the sequential correction of a larger control vector and bring spatial variability in the friction and input of water; thus improving water level in reanalysis and forecast. The frequent-in-time and sparse-in-space in-situ observations can be greatly complemented by 2D flood extent maps at Sentinel-1 overpass times to inform on the dynamics of the flow over the entire catchment. A great number of DA studies using RS-derived water levels for state and input correction are found in the literature and reported in \cite{grimaldi2016remote} and \cite{Dasgupta2021review}. Beyond these approaches; several assimilation strategies will be investigated in future works to benefit the most out of the RS-derived flood extent observations. Possible leads are the assimilation of the number of wet pixels within predefined subdomains, the assimilation of the position of the wet/dry interface, and the assimilation of the probability of flooding of a pixel. These approaches will imply the development of corresponding observation operators and the definition of the appropriate control vector. Finally, the general perspective for this work stands in the development of a rapid mapping tool for flood alert dedicated to decision support systems at global scale. This is a real challenge in terms of computational resources limitations, data acquisition and processing. 


%



\section*{Acknowledgment}

Funding for this work was provided by CNES and CERFACS.  
The authors gratefully thank the  Electricité de France (EDF) for providing the Telemac2D model on the Garonne Downstream catchment, and the SCHAPI, SPCs Garonne-Tarn-Lot and Gironde-Adour-Dordogne for providing in-situ data. They also would like to thank S. El Garroussi and M. De Lozzo for former developments on the Telemac System with API dedicated to data-driven simulations, and C. Delenne, S. Peña Luque, R. Hostache, M. Chini for fruitful discussions. 
Lastly, the authors would like to thank the anonymous reviewers whose comments and suggestions helped improve this manuscript.

\ifCLASSOPTIONcaptionsoff
  \newpage
\fi



\bibliographystyle{IEEEtran}
\bibliography{IEEEabrv,ref.bib}

 \newcommand{\noop}[1]{}
\begin{thebibliography}{10}
\providecommand{\url}[1]{#1}
\csname url@samestyle\endcsname
\providecommand{\newblock}{\relax}
\providecommand{\bibinfo}[2]{#2}
\providecommand{\BIBentrySTDinterwordspacing}{\spaceskip=0pt\relax}
\providecommand{\BIBentryALTinterwordstretchfactor}{4}
\providecommand{\BIBentryALTinterwordspacing}{\spaceskip=\fontdimen2\font plus
\BIBentryALTinterwordstretchfactor\fontdimen3\font minus
  \fontdimen4\font\relax}
\providecommand{\BIBforeignlanguage}[2]{{%
\expandafter\ifx\csname l@#1\endcsname\relax
\typeout{** WARNING: IEEEtran.bst: No hyphenation pattern has been}%
\typeout{** loaded for the language `#1'. Using the pattern for}%
\typeout{** the default language instead.}%
\else
\language=\csname l@#1\endcsname
\fi
#2}}
\providecommand{\BIBdecl}{\relax}
\BIBdecl

\bibitem{IPCC2021}
V.~Masson-Delmotte, P.~Zhai, A.~Pirani, S.~Connors, C.~Péan, S.~Berger,
  N.~Caud, Y.~Chen, L.~Goldfarb, M.~I. Gomis, M.~Huang, K.~Leitzell, E.~Lonnoy,
  J.~B.~R. matthewes, T.~K. Maycock, T.~Waterfield, O.~Yelekci, R.~Yu, and
  B.~e. Zhou, ``Ipcc 2021: Climate change 2021: The physical science basis.
  contribution of working group i to the sixth assessment report of the
  intergovernmental panel on climate change.'' \emph{Cambridge Univertisty
  Press. In press}, 2021.

\bibitem{IPCC2021SPM}
------, ``Ipcc 2021: Summary for policymakers. in: Climate change 2021: The
  physical science basis. contribution of working group i to the sixth
  assessment report of the intergovernmental panel on climate change.''
  \emph{Cambridge Univertisty Press. In press}, 2021.

\bibitem{odry2017comparison}
J.~Odry and P.~Arnaud, ``Comparison of flood frequency analysis methods for
  ungauged catchments in france,'' \emph{Geosciences}, vol.~7, no.~3, p.~88,
  2017.

\bibitem{kettig}
P.~Kettig, B.~Simon, G.~Blanchet, C.~Taillan, S.~Ricci, T.~H. Nguyen, T.~Huang,
  A.~Altinok, N.~T. Chung, G.~Valladeau, R.~Goeury, and A.~Roumagnac, ``The
  sco-flooddam project: New observing strate-gies for flood detection, alert
  and rapid mapping,'' in \emph{IGARSS 2021 - 2021 IEEE International
  Geoscience and Remote Sensing Symposium}, \noop{2021}in press.

\bibitem{swe2d}
J.~C. de~Saint-Venant, ``Th\'eorie du mouvement non-permanent des eaux, avec
  application aux crues des rivi\`eres et \`a l'introduction des mar\'ees dans
  leur lit,'' \emph{C. R. Acad. Sc. Paris}, vol.~73, pp. 147--154, 1871.

\bibitem{navierstokes}
H.~Sohr, \emph{The Navier-Stokes Equations}.\hskip 1em plus 0.5em minus
  0.4em\relax Birkhäuser Basel, 2001.

\bibitem{Bates1997}
P.~Bates, M.~Horritt, C.~Smith, and D.~Mason,
  ``\BIBforeignlanguage{English}{Integrating remote sensing observations of
  flood hydrology and hydraulic modelling},''
  \emph{\BIBforeignlanguage{English}{Hydrological Processes}}, vol.~11, no.~14,
  pp. 1777--1795, Nov. 1997.

\bibitem{Bates2004}
P.~Bates, ``\BIBforeignlanguage{English}{Remote sensing and flood inundation
  modelling},'' \emph{\BIBforeignlanguage{English}{Hydrological Processes}},
  vol.~18, pp. 2593 -- 2597, 2004, publisher: John Wiley and Sons.

\bibitem{Wood2016}
\BIBentryALTinterwordspacing
M.~Wood, R.~Hostache, J.~Neal, T.~Wagener, L.~Giustarini, M.~Chini, G.~Corato,
  P.~Matgen, and P.~Bates, ``Calibration of channel depth and friction
  parameters in the lisflood-fp hydraulic model using medium-resolution sar
  data and identifiability techniques,'' \emph{Hydrology and Earth System
  Sciences}, vol.~20, no.~12, pp. 4983--4997, 2016. [Online]. Available:
  \url{https://hess.copernicus.org/articles/20/4983/2016/}
\BIBentrySTDinterwordspacing

\bibitem{Horritt2000}
\BIBentryALTinterwordspacing
M.~S. Horritt, ``Calibration of a two-dimensional finite element flood flow
  model using satellite radar imagery,'' \emph{Water Resources Research},
  vol.~36, no.~11, pp. 3279--3291, 2000. [Online]. Available:
  \url{https://agupubs.onlinelibrary.wiley.com/doi/abs/10.1029/2000WR900206}
\BIBentrySTDinterwordspacing

\bibitem{Werner2005}
\BIBentryALTinterwordspacing
M.~Werner, S.~Blazkova, and J.~Peter, ``Spatially distributed observations in
  constraining inundation modelling uncertainties,'' \emph{Hydrological
  Processes}, vol.~19, no.~16, pp. 3081--3096, 2005. [Online]. Available:
  \url{https://onlinelibrary.wiley.com/doi/abs/10.1002/hyp.5833}
\BIBentrySTDinterwordspacing

\bibitem{mirouze2019impact}
I.~Mirouze, S.~Ricci, and N.~Goutal, ``The impact of observation spatial and
  temporal densification in an ensemble kalman filter,'' in \emph{XXVIth
  TELEMAC-MASCARET User Conference, 15th to 17th October 2019, Toulouse}, 2019.

\bibitem{ide1997unified}
K.~Ide, P.~Courtier, M.~Ghil, and A.~C. Lorenc, ``Unified notation for data
  assimilation: Operational, sequential and variational (gtspecial issueltdata
  assimilation in meteology and oceanography: Theory and practice),''
  \emph{Journal of the Meteorological Society of Japan. Ser. II}, vol.~75,
  no.~1B, pp. 181--189, 1997.

\bibitem{kalnay2003atmospheric}
E.~Kalnay, \emph{Atmospheric modeling, data assimilation and
  predictability}.\hskip 1em plus 0.5em minus 0.4em\relax Cambridge university
  press, 2003.

\bibitem{asch2016data}
M.~Asch, M.~Bocquet, and M.~Nodet, \emph{Data assimilation: methods,
  algorithms, and applications}.\hskip 1em plus 0.5em minus 0.4em\relax SIAM,
  2016.

\bibitem{liu2012advancing}
Y.~Liu, A.~Weerts, M.~Clark, H.-J. Hendricks~Franssen, S.~Kumar, H.~Moradkhani,
  D.-J. Seo, D.~Schwanenberg, P.~Smith, A.~Van~Dijk \emph{et~al.}, ``Advancing
  data assimilation in operational hydrologic forecasting: progresses,
  challenges, and emerging opportunities,'' \emph{Hydrology and Earth System
  Sciences}, vol.~16, no.~10, pp. 3863--3887, 2012.

\bibitem{evensen1994sequential}
G.~Evensen, ``Sequential data assimilation with a nonlinear quasi-geostrophic
  model using monte carlo methods to forecast error statistics,'' \emph{Journal
  of Geophysical Research: Oceans}, vol.~99, no.~C5, pp. 10\,143--10\,162,
  1994.

\bibitem{BARTHELEMY2017210}
\BIBentryALTinterwordspacing
S.~Barthélémy, S.~Ricci, M.~Rochoux, E.~{Le Pape}, and O.~Thual,
  ``Ensemble-based data assimilation for operational flood forecasting – on
  the merits of state estimation for 1d hydrodynamic forecasting through the
  example of the “adour maritime” river,'' \emph{Journal of Hydrology},
  vol. 552, pp. 210--224, 2017. [Online]. Available:
  \url{https://www.sciencedirect.com/science/article/pii/S0022169417304201}
\BIBentrySTDinterwordspacing

\bibitem{ad2001global}
A.~H. Group, C.~V{\"o}r{\"o}smarty, A.~Askew, W.~Grabs, R.~Barry, C.~Birkett,
  P.~D{\"o}ll, B.~Goodison, A.~Hall, R.~Jenne \emph{et~al.}, ``Global water
  data: A newly endangered species,'' \emph{Eos, Transactions American
  Geophysical Union}, vol.~82, no.~5, pp. 54--58, 2001.

\bibitem{Jafarzadegan2019}
K.~Jafarzadegan, P.~Abbaszadeh, and H.~Moradkhani, ``Sequential data
  assimilation for real-time probabilistic flood inundation mapping.''
  \emph{Hydrol. Earth Syst. Sci.}, vol.~25, pp. 4995--5011, 2021.

\bibitem{Schumann2009}
\BIBentryALTinterwordspacing
G.~Schumann, P.~D. Bates, M.~S. Horritt, P.~Matgen, and F.~Pappenberger,
  ``Progress in integration of remote sensing–derived flood extent and stage
  data and hydraulic models,'' \emph{Reviews of Geophysics}, vol.~47, no.~4,
  2009. [Online]. Available:
  \url{https://agupubs.onlinelibrary.wiley.com/doi/abs/10.1029/2008RG000274}
\BIBentrySTDinterwordspacing

\bibitem{hostache2009water}
R.~Hostache, P.~Matgen, G.~Schumann, C.~Puech, L.~Hoffmann, and L.~Pfister,
  ``Water level estimation and reduction of hydraulic model calibration
  uncertainties using satellite sar images of floods,'' \emph{IEEE Transactions
  on Geoscience and Remote Sensing}, vol.~47, no.~2, pp. 431--441, 2009.

\bibitem{henderson1998principles}
\BIBentryALTinterwordspacing
F.~M. Henderson and A.~J. Lewis, ``Principles and applications of imaging
  radar. manual of remote sensing: Third edition, volume 2,'' 12 1998.
  [Online]. Available: \url{https://www.osti.gov/biblio/293027}
\BIBentrySTDinterwordspacing

\bibitem{chini2017hierarchical}
M.~Chini, R.~Hostache, L.~Giustarini, and P.~Matgen, ``A hierarchical
  split-based approach for parametric thresholding of sar images: Flood
  inundation as a test case,'' \emph{IEEE Transactions on Geoscience and Remote
  Sensing}, vol.~55, no.~12, pp. 6975--6988, 2017.

\bibitem{cian2018flood}
F.~Cian, M.~Marconcini, P.~Ceccato, and C.~Giupponi, ``Flood depth estimation
  by means of high-resolution sar images and lidar data,'' \emph{Natural
  Hazards and Earth System Sciences}, vol.~18, no.~11, pp. 3063--3084, 2018.

\bibitem{cian2018normalized}
F.~Cian, M.~Marconcini, and P.~Ceccato, ``Normalized difference flood index for
  rapid flood mapping: Taking advantage of eo big data,'' \emph{Remote Sensing
  of Environment}, vol. 209, pp. 712--730, 2018.

\bibitem{rattich2020automatic}
M.~R{\"a}ttich, S.~Martinis, and M.~Wieland, ``Automatic flood duration
  estimation based on multi-sensor satellite data,'' \emph{Remote Sensing},
  vol.~12, no.~4, p. 643, 2020.

\bibitem{Jung2012}
\BIBentryALTinterwordspacing
H.~C. Jung, M.~Jasinski, J.-W. Kim, C.~K. Shum, P.~Bates, J.~Neal, H.~Lee, and
  D.~Alsdorf, ``Calibration of two-dimensional floodplain modeling in the
  central atchafalaya basin floodway system using sar interferometry,''
  \emph{Water Resources Research}, vol.~48, no.~7, 2012. [Online]. Available:
  \url{https://agupubs.onlinelibrary.wiley.com/doi/abs/10.1029/2012WR011951}
\BIBentrySTDinterwordspacing

\bibitem{grimaldi2016remote}
S.~Grimaldi, Y.~Li, V.~R. Pauwels, and J.~P. Walker, ``Remote sensing-derived
  water extent and level to constrain hydraulic flood forecasting models:
  Opportunities and challenges,'' \emph{Surveys in Geophysics}, vol.~37, no.~5,
  pp. 977--1034, 2016.

\bibitem{Schumann2012}
\BIBentryALTinterwordspacing
G.~J.-P. Schumann, P.~D. Bates, G.~Di~Baldassarre, and D.~C. Mason, \emph{The
  Use of Radar Imagery in Riverine Flood Inundation Studies}.\hskip 1em plus
  0.5em minus 0.4em\relax John Wiley and Sons, Ltd, 2012, ch.~6, pp. 115--140.
  [Online]. Available:
  \url{https://onlinelibrary.wiley.com/doi/abs/10.1002/9781119940791.ch6}
\BIBentrySTDinterwordspacing

\bibitem{Dasgupta2021review}
\BIBentryALTinterwordspacing
A.~Dasgupta, R.~Hostache, R.~Ramsankaran, S.~Grimaldi, P.~Matgen, M.~Chini,
  V.~R. Pauwels, and J.~P. Walker, ``Chapter 12 - earth observation and
  hydraulic data assimilation for improved flood inundation forecasting,'' in
  \emph{Earth Observation for Flood Applications}, ser. Earth Observation,
  G.~J.-P. Schumann, Ed.\hskip 1em plus 0.5em minus 0.4em\relax Elsevier, 2021,
  pp. 255--294. [Online]. Available:
  \url{https://www.sciencedirect.com/science/article/pii/B9780128194126000122}
\BIBentrySTDinterwordspacing

\bibitem{mason2012automatic}
D.~Mason, G.-P. Schumann, J.~Neal, J.~Garcia-Pintado, and P.~Bates, ``Automatic
  near real-time selection of flood water levels from high resolution synthetic
  aperture radar images for assimilation into hydraulic models: A case study,''
  \emph{Remote Sensing of Environment}, vol. 124, pp. 705--716, 2012.

\bibitem{giustarini2011assimilating}
L.~Giustarini, P.~Matgen, R.~Hostache, M.~Montanari, D.~Plaza, V.~Pauwels,
  G.~De~Lannoy, R.~D. Keyser, L.~Pfister, L.~Hoffmann \emph{et~al.},
  ``Assimilating sar-derived water level data into a hydraulic model: a case
  study,'' \emph{Hydrology and Earth System Sciences}, vol.~15, no.~7, pp.
  2349--2365, 2011.

\bibitem{scarpino2018multitemporal}
S.~Scarpino, R.~Albano, A.~Cantisani, L.~Mancusi, A.~Sole, and G.~Milillo,
  ``Multitemporal sar data and 2d hydrodynamic model flood scenario dynamics
  assessment,'' \emph{ISPRS International Journal of Geo-Information}, vol.~7,
  no.~3, p. 105, 2018.

\bibitem{grimaldi2018effective}
S.~Grimaldi, Y.~Li, J.~Walker, and V.~Pauwels, ``Effective representation of
  river geometry in hydraulic flood forecast models,'' \emph{Water Resources
  Research}, vol.~54, no.~2, pp. 1031--1057, 2018.

\bibitem{hostache2018near}
R.~Hostache, M.~Chini, L.~Giustarini, J.~Neal, D.~Kavetski, M.~Wood, G.~Corato,
  R.-M. Pelich, and P.~Matgen, ``Near-real-time assimilation of sar-derived
  flood maps for improving flood forecasts,'' \emph{Water Resources Research},
  vol.~54, no.~8, pp. 5516--5535, 2018.

\bibitem{DiMauro2021}
C.~Di~Mauro, R.~Hostache, P.~Matgen, R.~Pelich, M.~Chini, P.~Jan~van Leeuwen,
  N.~Nichols, and G.~Bloschl, ``Assimilation of probabilistic flood maps from
  sar data into a coupled hydrologic-hydraulic forecasting model: a proof of
  concept.'' \emph{Hydrol. Earth Syst. Sci.}, vol.~25, pp. 4081--4097, 2021.

\bibitem{Dasgupta2020}
A.~Dasgupta, R.~Hostache, R.~Ramsankaran, G.~J.-P. Schumann, S.~Grimaldi,
  V.~R.~N. Pauwels, and J.~P. Walker, ``A mutual information-based likelihood
  function for particle filter flood extent assimilation,'' \emph{Water
  Resources Research}, vol.~57, no.~2, p. e2020WR027859, 2021.

\bibitem{Dasgupta2021network}
------, ``On the impacts of observation location, timing, and frequency on
  flood extent assimilation performance,'' \emph{Water Resources Research},
  vol.~57, no.~2, p. e2020WR028238, 2021.

\bibitem{matgen2011towards}
P.~Matgen, R.~Hostache, G.~Schumann, L.~Pfister, L.~Hoffmann, and H.~Savenije,
  ``Towards an automated sar-based flood monitoring system: Lessons learned
  from two case studies,'' \emph{Physics and Chemistry of the Earth, Parts
  A/B/C}, vol.~36, no. 7-8, pp. 241--252, 2011.

\bibitem{Giustarini2013}
L.~Giustarini, R.~Hostache, P.~Matgen, G.~J.-P. Schumann, P.~D. Bates, and
  D.~C. Mason, ``A change detection approach to flood mapping in urban areas
  using terrasar-x,'' \emph{IEEE Transactions on Geoscience and Remote
  Sensing}, vol.~51, no.~4, pp. 2417--2430, 2013.

\bibitem{Chini2016}
M.~Chini, A.~Papastergios, L.~Pulvirenti, N.~Pierdicca, P.~Matgen, and
  I.~Parcharidis, ``Sar coherence and polarimetric information for improving
  flood mapping,'' in \emph{2016 IEEE International Geoscience and Remote
  Sensing Symposium (IGARSS)}, 2016, pp. 7577--7580.

\bibitem{wagener2003towards}
T.~Wagener, N.~McIntyre, M.~Lees, H.~Wheater, and H.~Gupta, ``Towards reduced
  uncertainty in conceptual rainfall-runoff modelling: Dynamic identifiability
  analysis,'' \emph{Hydrological processes}, vol.~17, no.~2, pp. 455--476,
  2003.

\bibitem{neal2012simple}
J.~Neal, G.~Schumann, and P.~D. Bates, ``A simple model for simulating river
  hydraulics and floodplain inundation over large and data sparse areas,''
  \emph{Water Resour Res}, vol.~48, 2012.

\bibitem{cooper2019observation}
E.~S. Cooper, S.~L. Dance, J.~Garc{\'\i}a-Pintado, N.~K. Nichols, and P.~J.
  Smith, ``Observation operators for assimilation of satellite observations in
  fluvial inundation forecasting,'' \emph{Hydrology and Earth System Sciences},
  vol.~23, no.~6, pp. 2541--2559, 2019.

\bibitem{gauckler1867etudes}
P.~Gauckler, \emph{Etudes Th{\'e}oriques et Pratiques sur l'Ecoulement et le
  Mouvement des Eaux}.\hskip 1em plus 0.5em minus 0.4em\relax Gauthier-Villars,
  1867.

\bibitem{hervouet2007hydrodynamics}
J.-M. Hervouet, \emph{Hydrodynamics of free surface flows: modelling with the
  finite element method}.\hskip 1em plus 0.5em minus 0.4em\relax Wiley Online
  Library, 2007, vol. 360.

\bibitem{besnard2011comparaison}
A.~Besnard and N.~Goutal, ``Comparaison de mod{\`e}les 1d {\`a} casiers et 2d
  pour la mod{\'e}lisation hydraulique d’une plaine d’inondation--cas de la
  garonne entre tonneins et la r{\'e}ole,'' \emph{La Houille Blanche}, no.~3,
  pp. 42--47, 2011.

\bibitem{Horritt2002}
M.~Horritt and P.~Bates, ``\BIBforeignlanguage{English}{Evaluation of 1-d and
  2-d models for predicting river flood inundation},''
  \emph{\BIBforeignlanguage{English}{Journal of Hydrology}}, vol. 180, pp. 87
  -- 99, 2002.

\bibitem{Dibaldassarre2009}
\BIBentryALTinterwordspacing
G.~Di~Baldassarre and A.~Montanari, ``Uncertainty in river discharge
  observations: a quantitative analysis,'' \emph{Hydrology and Earth System
  Sciences}, vol.~13, no.~6, pp. 913--921, 2009. [Online]. Available:
  \url{https://hess.copernicus.org/articles/13/913/2009/}
\BIBentrySTDinterwordspacing

\bibitem{ricci2011correction}
S.~Ricci, A.~Piacentini, O.~Thual, E.~L. Pape, and G.~Jonville, ``Correction of
  upstream flow and hydraulic state with data assimilation in the context of
  flood forecasting,'' \emph{Hydrology and Earth System Sciences}, vol.~15,
  no.~11, pp. 3555--3575, 2011.

\bibitem{Madsen2005}
H.~Madsen and C.~Skotner, ``Adaptive state updating in real-time river flow
  forecasting - a combined filtering and error forecasting procedure,''
  \emph{Journal of Hydrology}, vol. 308, pp. 302--312, 2005.

\bibitem{Neal2007}
J.~Neal and C.~Jeffrey, ``Flood inundation model updating using an ensemble
  kalman filter and spatially distributed measurements,'' \emph{Journal of
  Hydrology}, vol. 336, pp. 401--415, 2007.

\bibitem{Neal2009}
J.~Neal, C.~Jeffrey, P.~Atkinson, and C.~Hutton, ``Evaluating the utility of
  the ensemble transform kalman filter for adaptive sampling when updating a
  hydrodynamic model,'' \emph{Journal of Hydrology}, vol. 375, no. 3-4, pp.
  589--600, 2009.

\bibitem{torres2012gmes}
R.~Torres, P.~Snoeij, D.~Geudtner, D.~Bibby, M.~Davidson, E.~Attema, P.~Potin,
  B.~Rommen, N.~Floury, M.~Brown \emph{et~al.}, ``Gmes sentinel-1 mission,''
  \emph{Remote Sensing of Environment}, vol. 120, pp. 9--24, 2012.

\bibitem{2020AGUFMIN041..09H}
T.~{Huang}, S.~{Baillarin}, A.~{Altinok}, G.~{Blanchet}, J.~{Hausman},
  P.~{Kettig}, and S.~{Shah}, ``{Distributed Machine Learning and Data Fusion
  for Flood Detection and Monitoring},'' in \emph{AGU Fall Meeting Abstracts},
  vol. 2020, Dec. 2020, pp. IN041--09.

\bibitem{pal2005random}
M.~Pal, ``Random forest classifier for remote sensing classification,''
  \emph{International journal of remote sensing}, vol.~26, no.~1, pp. 217--222,
  2005.

\bibitem{belgiu2016random}
M.~Belgiu and L.~Dr{\u{a}}gu{\c{t}}, ``Random forest in remote sensing: A
  review of applications and future directions,'' \emph{ISPRS journal of
  photogrammetry and remote sensing}, vol. 114, pp. 24--31, 2016.

\bibitem{pekel2016high}
J.-F. Pekel, A.~Cottam, N.~Gorelick, and A.~S. Belward, ``High-resolution
  mapping of global surface water and its long-term changes,'' \emph{Nature},
  vol. 540, no. 7633, pp. 418--422, 2016.

\bibitem{yamazaki2017high}
D.~Yamazaki, D.~Ikeshima, R.~Tawatari, T.~Yamaguchi, F.~O'Loughlin, J.~C. Neal,
  C.~C. Sampson, S.~Kanae, and P.~D. Bates, ``A high-accuracy map of global
  terrain elevations,'' \emph{Geophysical Research Letters}, vol.~44, no.~11,
  pp. 5844--5853, 2017.

\bibitem{ronneberger2015u}
O.~Ronneberger, P.~Fischer, and T.~Brox, ``U-net: Convolutional networks for
  biomedical image segmentation,'' in \emph{International Conference on Medical
  image computing and computer-assisted intervention}.\hskip 1em plus 0.5em
  minus 0.4em\relax Springer, 2015, pp. 234--241.

\bibitem{rs12030513}
\BIBentryALTinterwordspacing
D.~Derksen, J.~Inglada, and J.~Michel, ``Geometry aware evaluation of
  handcrafted superpixel-based features and convolutional neural networks for
  land cover mapping using satellite imagery,'' \emph{Remote Sensing}, vol.~12,
  no.~3, 2020. [Online]. Available:
  \url{https://www.mdpi.com/2072-4292/12/3/513}
\BIBentrySTDinterwordspacing

\bibitem{grimaldi2020}
\BIBentryALTinterwordspacing
S.~Grimaldi, J.~Xu, Y.~Li, V.~Pauwels, and J.~Walker, ``Flood mapping under
  vegetation using single sar acquisitions,'' \emph{Remote Sensing of
  Environment}, vol. 237, p. 111582, 2020. [Online]. Available:
  \url{https://www.sciencedirect.com/science/article/pii/S0034425719306029}
\BIBentrySTDinterwordspacing

\bibitem{stephens2014problems}
E.~Stephens, G.~Schumann, and P.~Bates, ``Problems with binary pattern measures
  for flood model evaluation,'' \emph{Hydrological Processes}, vol.~28, no.~18,
  pp. 4928--4937, 2014.

\bibitem{NguyenTUC2021}
T.~H. Nguyen, A.~Delmotte, C.~Fatras, P.~Kettig, A.~Piacentini, and S.~Ricci,
  ``Validation and improvement of data assimilation for flood hydrodynamic
  modelling using sar imagery data,'' \emph{arXiv preprint arXiv:2109.07470},
  2021.

\bibitem{Schumann2008a}
G.~Schumann, P.~Matgen, and F.~Pappenberger, ``Conditioning water stages from
  satellite imagery on uncertain data points,'' \emph{IEEE Geoscience and
  Remote Sensing Letters}, vol.~5, no.~4, pp. 810--813, 2008.

\bibitem{Schumann2008b}
G.~Schumann, F.~Pappenberger, and P.~Matgen, ``Estimating uncertainty
  associated with water stages from a single sar image,'' \emph{Advances in
  Water Resources}, vol.~31, no.~8, pp. 1038--1047, 2008.

\bibitem{Blosch2001}
G.~Blosch, ``Scaling in hydrology,'' \emph{Hydrological processess}, vol.~15,
  pp. 709--711, 2001.

\bibitem{Wesemael2018}
\BIBentryALTinterwordspacing
A.~{Van Wesemael}, L.~Landuyt, H.~Lievens, and N.~E. Verhoest, ``Improving
  flood inundation forecasts through the assimilation of in situ floodplain
  water level measurements based on alternative observation network
  configurations,'' \emph{Advances in Water Resources}, vol. 130, pp. 229--243,
  2019. [Online]. Available:
  \url{https://www.sciencedirect.com/science/article/pii/S0309170819300752}
\BIBentrySTDinterwordspacing

\bibitem{Ziliani2019}
\BIBentryALTinterwordspacing
M.~G. Ziliani, R.~Ghostine, B.~Ait-El-Fquih, M.~F. McCabe, and I.~Hoteit,
  ``Enhanced flood forecasting through ensemble data assimilation and joint
  state-parameter estimation,'' \emph{Journal of Hydrology}, vol. 577, p.
  123924, 2019. [Online]. Available:
  \url{https://www.sciencedirect.com/science/article/pii/S0022169419306444}
\BIBentrySTDinterwordspacing

\end{thebibliography}
\end{document}